%% file: main.tex
\documentclass{nature}

\usepackage{float}
\usepackage[labelfont=bf]{caption}
\usepackage{graphicx}
\makeatletter
\let\saved@includegraphics\includegraphics
\AtBeginDocument{\let\includegraphics\saved@includegraphics}
\renewenvironment{figure}{\@float{figure}}
{\end@float}
\makeatother
\usepackage{xcolor}
\usepackage{amsmath}
\usepackage{comment}
\usepackage[%
]{hyperref}

\newcommand{\trR}{$\Delta R/R$}
\newcommand{\dtpp}{$\Delta t_{pp}$}

\title{Coherent control through phonon anharmonicity}

\author{Gili Scharf$^1$, Tomer Hasharoni$^1$, Lara Donval$^1$, Leah Ben Gur$^1$, Alon Ron$^1$}

\begin{document}

\maketitle

\begin{affiliations}

 \item Raymond and Beverly Sackler School of Physics and Astronomy, Tel-Aviv 69978, Israel
\end{affiliations}

\begin{abstract}
Anharmonic lattice vibrations play a key role in many physical phenomena. They govern the heat conductivity of solids, strongly affect the phonon spectra, play a prominent role in soft mode phase transitions, allow ultrafast engineering of material properties and more. The most direct evidence for anharmonicity is to measure the oscillation frequency changing as a function of the oscillation amplitude. For lattice vibrations, this is not a trivial task, and anharmonicity is probed indirectly through its effects on thermodynamic properties and spectral features or through coherent decay of one mode to another. However, measurement and control of the anharmonicity of a single Raman mode is still lacking. We show that ultrafast double pump-probe spectroscopy could be used to directly observe frequency shifts of Raman phonons as a function of the oscillation amplitude and disentangle the coherent contributions from quasi-harmonic sources such as temperature and changes to the carrier density in the thermoelectric materials SnTe and SnSe.  Moreover, we show that coherent displacive phononic excitations in tandem with electron-phonon coupling is a pathway to dynamically control phonon anharmonicity. Our results have dramatic implications for the material engineering of future thermoelectrics. Moreover, our methodology could be used to isolate the basic mechanisms driving optically induced phase transitions and other nonlinear phenomena based on their unique timestamps.

\end{abstract}

    The properties of phonons and their interplay with other constituents of a solid play a crucial role in determining material properties. One such property is the phonon anharmonicity, which determines the lattice thermal conductivity, and understanding it is critical for the development and engineering of new thermoelectric and energy materials\cite{delaire2011giant,wei2021phonon}. Phonon anharmonicity could be indirectly probed by measuring the change in scattering or thermodynamic properties as a function of external stimuli such as temperature and pressure. For example, it can be observed as line-width broadening and changes to the phonon frequency, as well as variations in thermal conductivity or expansion\cite{knoop2023anharmonicity,wallace1965thermal} as the temperature changes. Alternatively, its effects on the phonon spectrum could be probed through an inelastic scattering of neutrons\cite{delaire2011giant} and X-rays\cite{budai2014metallization}. Ultrafast pump-probe spectroscopy has also been used to observe anharmonic coupling between phonon modes. For example, Teitelbaum \textit{et al.}\cite{teitelbaum2018direct} showed that the $A_{1g}$~phonon in excited bismuth decays to pairs of acoustic modes with opposite momenta, which drove temporal oscillations at the corresponding frequency.
    Similarly, off-diagonal peaks in two-dimensional spectroscopy measure the coupling between different vibrational modes of molecules and solids\cite{golonzka2001coupling,fulmer2004pulse,golonzka2001vibrational,lin2022mapping}. Another example is the nonlinear coupling between infrared-active modes and symmetric Raman modes was observed by F{\"o}rst \textit{et al.}\cite{forst2011nonlinear}, pioneering the field of nonlinear phononics. The common feature of all of these experiments is that they probed anharmonic coupling between different vibrational modes. However, to our knowledge, no experiment has yet measured the frequency shifts of a single mode resulting from variations in the oscillation amplitude.
    \par

    Anharmonicity is particularly important in thermoelectric materials and tuning it, as well as the frequencies, lifetime, and dispersion of both acoustic and optical modes are utilized as control knobs for thermoelectric efficiency. As an example, Toberer \textit{et al.} show that phonon scattering times from Umklapp and point defects scale as $\tau_U\sim\omega^{-2}$~and $\tau_{PD}\sim\omega^{-4}$~respectively\cite{toberer2011phonon}. This dependence of phonon scattering times on frequencies is directly related to the phonon thermal conductivity and thus affects their thermoelectric performance. We chose to work with SnTe and SnSe. SnTe is a promising thermoelectric material\cite{brebrick1963anomalous}, whose anharmonicity is currently a topic of extensive study\cite{li2014phonon,wei2021phonon,lee2014resonant,ribeiro2018strong}. Additionally, SnTe has recently gained further attention thanks to its topological crystalline insulator properties\cite{hsieh2012topological}. SnSe could be thought of as a lower symmetry, orthorhombic variant of the rock-salt NaCl-like structure of SnTe. In this work, we directly measure the anharmonicity of individual phonon modes by utilizing double pump-probe spectroscopy, a technique used in the past to control phase transitions using light\cite{johnson2024all,horstmann2020coherent,maklar2023coherent}, to coherently control phonon oscillations\cite{teitelbaum2018direct,hu2018femtosecond,sasaki2018coherent}, and more. Our findings reveal significant frequency shifts resulting from variations in oscillation amplitude. This method allows us to disentangle and quantify the contributions of different physical effects to ultrafast measurements using their temporal fingerprint, such as thermal, electronic, and anharmonic effects on phonon behavior.
\par

\begin{figure}[htbp]
    \centering
    \includegraphics[width=0.95\textwidth]{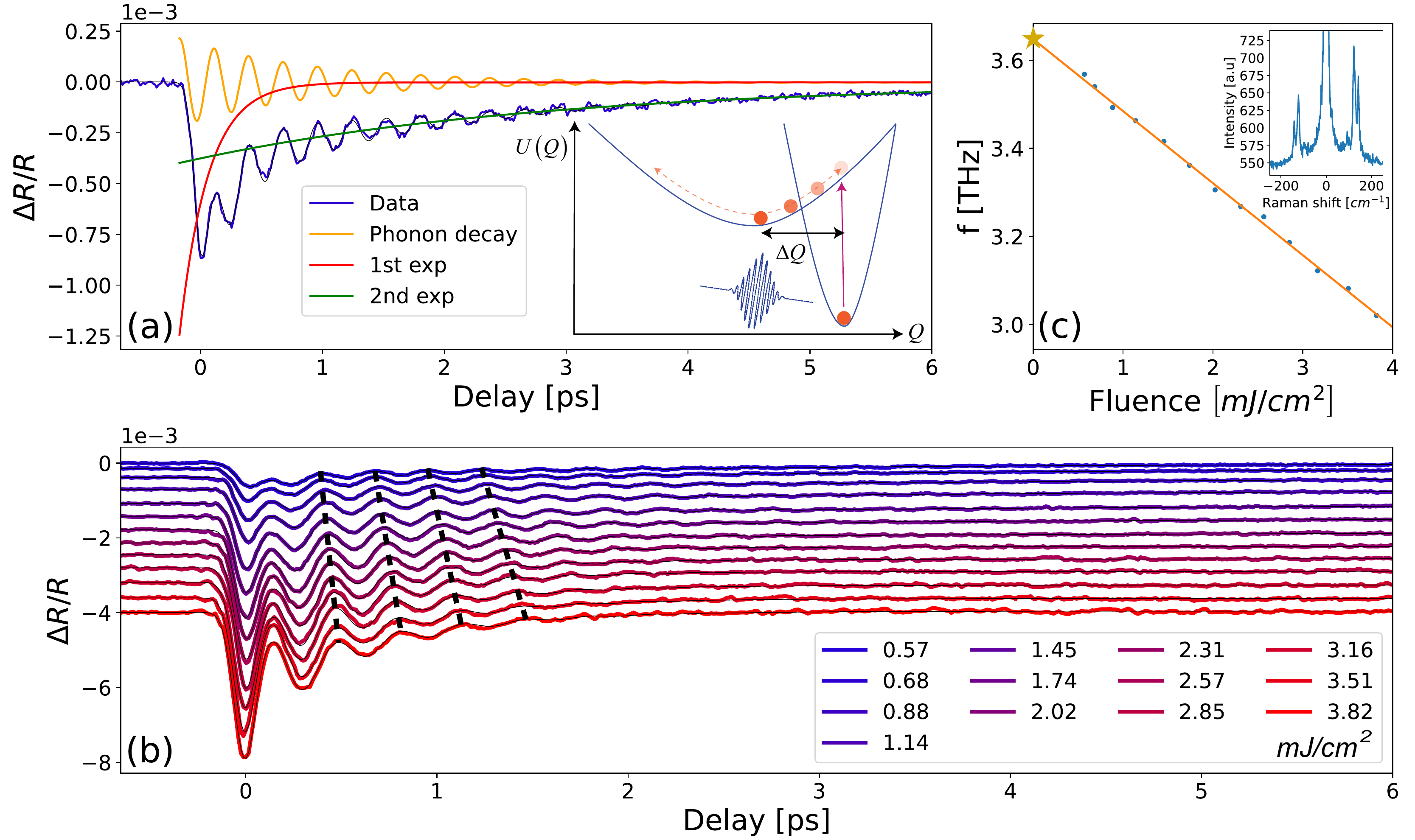}
    \caption{(a) \trR~of SnTe with an incident pump fluence of 0.684 $[mJ/cm^2]$. The blue line is the data, and the black curve is its fit. The red, green, and yellow solid lines are the decomposition of the fit to its exponential and oscillatory components. The inset illustrates the DECP process. As a result of the optical excitation the potential energy landscape $U(Q)$~for the phonons is displaced along the phonon coordinate $Q$~which initiates the oscillatory motion of the phonon (b) Fluence dependence of the \trR. Data is shown in color, and fits are shown as thin black lines. The dashed black lines are guides to the eye through the oscillatory peaks. Data is vertically shifted for clarity. Incident fluences are specified in units of $mJ/cm^2$. (c) Frequencies extracted from the fits to the data shown in panel (b) as a function of the applied pump fluence. The star marks the frequency measured by Raman scattering. In the inset, Raman scattering measurement. A star marks the stokes and anti-stokes peaks corresponding to the same mode measured using the pump-probe technique.}
    \label{Figure 1}
\end{figure}

    To excite coherent oscillations of the $A_g$~mode, we utilize the "displacive excitation of coherent phonons" (DECP) process\cite{vidal1992displacive}. Ultrashort laser pulses (70 and 126 fs) at 750 nm, resonant with a transition between  Te and Sn bands, are incident on the sample, resulting in photo-excitation of electrons from the Te anion to the Sn cation. This reduces the charge difference between the ions, resulting in a softening of the chemical bond between them. The charge transfer shifts the equilibrium position of the ions closer to one another, causing them to oscillate around the new equilibrium positions. 
    The inset to Figure \ref{Figure 1}(a) illustrates the DECP process for a harmonic potential. The red circle at the bottom curve depicts the phonon in the ground state. After the photoexcitation, the potential energy landscape shifts along the phonon coordinate by $\Delta Q$ and the phonon starts oscillating about the new minimum.
    \par

Figure\ref{Figure 1}(a) shows the relative change of sample reflectivity \trR~as a function of the time delay from a single pump excitation with an incident fluence of 0.68$[mJ/cm^2]$ (which translates to excitation density of $\approx 7.10\times10^{20} [cm^{-3}]$\cite{Suzuki_1995}). The signal could be decomposed to a decaying bi-exponential background overlayed with a damped oscillatory component. The colored solid lines in Figure\ref{Figure 1}(a) show the decomposition of the signal to its different components. We note that the oscillatory component displays a cosine-like behavior in agreement with the DECP mechanism\cite{vidal1992displacive}. The sign of the oscillatory component near time zero is positive. Understanding that the DECP mechanism in SnTe is initiated by bond softening enables us to use the positive sign of \trR~at time zero as a calibration, linking bond softening and the positive sign of the oscillatory component of \trR. The fast exponential component can be associated with electrons decaying back to their ground state due to the similar timescales observed using time-resolved photoemission\cite{ito2020observation}. The slower exponential component can be associated with thermal relaxation effects between the electronic and phononic baths. A longer time constant related to thermal relaxation effects of the phononic bath, is manifested on the short timescales of these measurements as a constant. Estimates of the expected time constants from known thermal conductivity and heat capacity of SnTe are shown in supplementary section 3. Figure\ref{Figure 1}(b) shows the fluence dependence of the \trR~signal of SnTe. The most striking aspect of the data is the redshift the oscillations exhibit towards higher fluences, indicating strong softening of the mode as the fluence is increased. Dashed lines are used as a guide to the eye to mark the shift in the oscillation peaks as the fluence is varied. The shifts in peak positions are very small for the zero and first peaks and increase for the later peaks, serving as a signature that the shifts originate from changes in the oscillation frequency rather than its phase. Figure\ref{Figure 1}(c) shows the frequency that was extracted by fitting the data in Figure\ref{Figure 1}(b) as a function of the applied fluence. The magnitude of the phonon softening at the highest fluence is almost 20\% lower than its value in equilibrium. An extrapolation of the oscillation frequency to zero fluence gives a frequency of $\mathrm{3.65\ THz}$ corresponding to 122 cm$^-1$, compatible with the known Ag phonon of SnTe\cite{su2023orietation} and our Raman measurements (inset), marked by a yellow star.

\par
 Optical phonon softening as a result of increasing laser fluence was observed in several compounds in the past\cite{huang2022observation,hase2002dynamics,fritz2007ultrafast,murray2005effect,he2016coherent}. The most relevant example for our work is SnSe, where the phonon softening was associated with softening of the $Se4p_x-Sn5s$~bond due to the photo-excitation of electrons\cite{huang2022observation}. Similar phenomena were also observed in bismuth, where the nature of the phonon softening, whether electronic or anharmonic, was debated in a series of papers\cite{hase2002dynamics,fritz2007ultrafast,murray2005effect}.
From these examples alone, it is clear that phonon softening as a function of the laser fluence can originate from various sources. This is due to the multitude of effects that occur when the pump pulse interacts with the sample, including heating, photo-excitation of electrons, and the excitation of coherent phonons. Ideally, one would isolate the contributions of the different effects by tuning them individually. While this is possible for infrared active modes via resonant excitation\cite{katayama2012ferroelectric,von2018probing,melnikov2023anharmonic,xu2025time} it is a very challenging task for the amplitude Raman active modes since their excitation mechanism is typically a non-resonant process such as DECP\cite{vidal1992displacive} which inherently involves an electronic excitation or a sudden temperature increase. Two unique cases are the experiments by Maehrlein \textit{et al.}\cite{maehrlein2017terahertz} and Giorgianni \textit{et al.}\cite{giorgianni2022terahertz} who were able to resonantly pump Raman active phonons in diamond and $ V_2O_3$, respectively, by using sum frequency of two low energy phonons in the mid-infrared and THz regimes, respectively. In the case of diamond, this was made possible thanks to the very high frequency (40 THz) of the particular mode and its unusually high Raman activity. For the case of $V_2O_3$, it was made possible thanks to a very strong THz field and similarly strong Raman activity. A double pump experiment by Murry \textit{et al.}\cite{murray2005effect} shows that in bismuth fluence dependent softening originates from the coupling of the $A_{1g}$~phonon to the electron-hole plasma by showing that the phonon frequency does not change as a function of the pump-pump delay. In our experiment, we go beyond what was previously demonstrated and show that the distinct transient characteristics of structural, electronic, and thermal effects serve as unique fingerprints to isolate and quantify their impacts on phonon anharmonicity using double pump-probe spectroscopy. This allows us to extract the mode-specific electron-phonon coupling and is made possible through our new modulation scheme described in the following paragraph.
 \par

\begin{figure}[!h]
    \centering
    \includegraphics[width=0.95\textwidth]{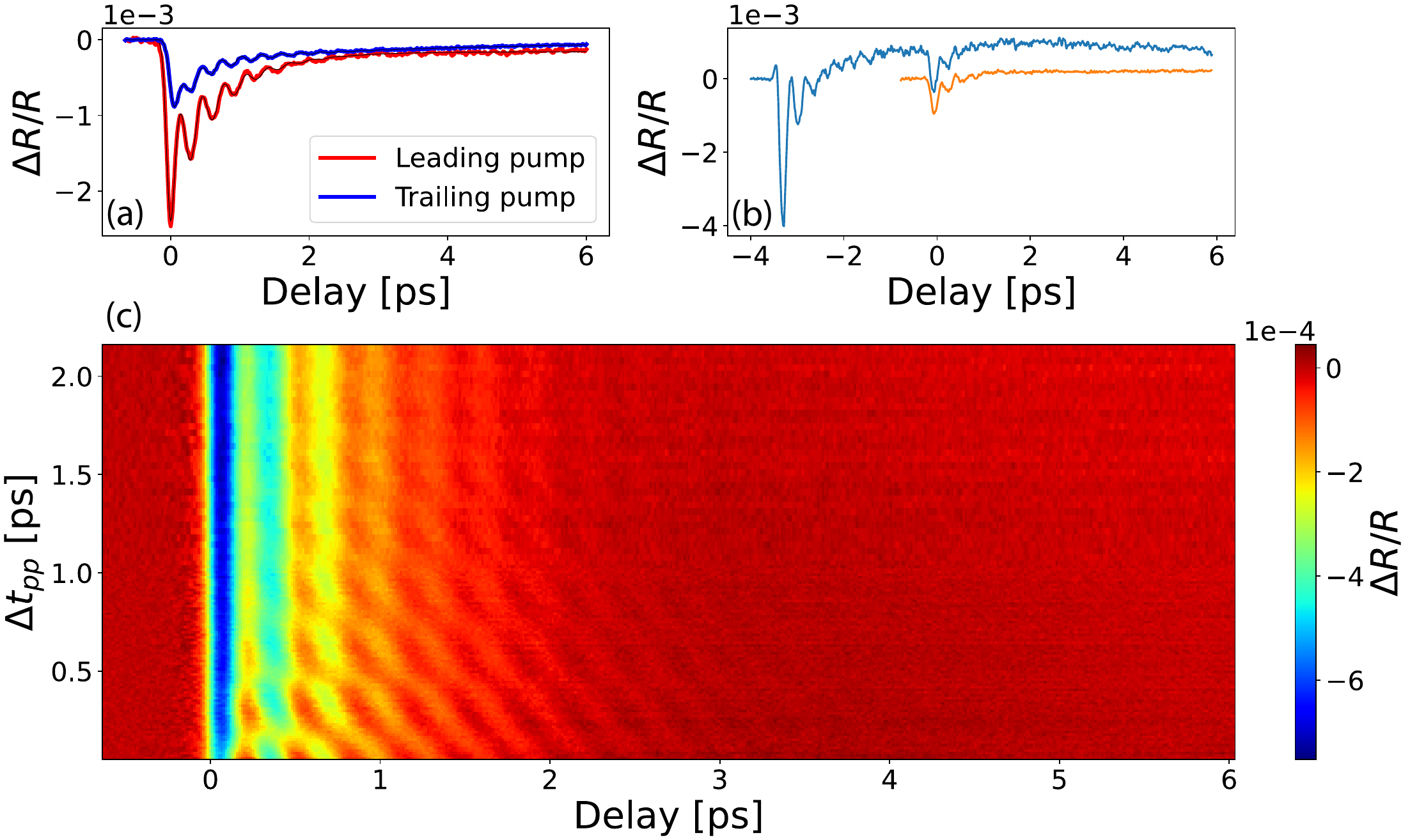}
    \caption{(a) \trR~data taken with single pump excitation. The blue curve is taken only with the trailing pump whereas the red curve is taken only with the leading pump. (b) \trR~data taken with both pumps incident on the sample with \dtpp~= 3.31[ps]. The blue curve was taken with the chopper modulating both pump beams and the orange curve corresponds to a measurement in which only the trailing pump is modulated by the chopper. (c) Color map of the double pump \trR~data. Colors represent the magnitude of \trR, the horizontal axis is the time delay between the probe and the trailing pump whereas the vertical axis is \dtpp.}
    \label{Figure 2}
\end{figure}

The incident fluences of the two pump beams were set to 2.822$[mJ/cm^2]$~for the leading pump and 0.571$[mJ/cm^2]$~for the trailing pump. These correspond to excitation densities of $2.929\times10^{21}$ and $5.93\times10^{20} [cm^{-3}]$ respectively. Their respective single pump \trR~traces are shown in Figure \ref{Figure 2}(a). The observed frequencies are 3.14 THz and 3.48 THz, corresponding to the pulses that will act as leading and trailing pump pulses in the double pump-probe experiment. Our optical setup allows us to isolate the portion of \trR~originating from the trailing pump by modulating it with a mechanical chopper. The efficiency of the isolation process is demonstrated in Figure \ref{Figure 2}(b). The time delay between the two pump pulses (\dtpp) was set to 3.31 ps, and the arrival time of the probe pulse was scanned relative to the two pump pulses. The blue trace shows the measurement when the chopper modulates both pump pulses, whereas the orange trace shows the measurement where only the trailing pump was. The measurement was repeated for different values of \dtpp~and the results are shown in Figure \ref{Figure 2}(c). The oscillation frequency of \trR~changes with \dtpp, as could be seen by the horizontal shift of the oscillation peaks. However, in contrast to the monotonically increasing mode softening, which was observed as a function of laser fluence (Figure \ref{Figure 1}), a nonmonotonic oscillatory behavior as a function of \dtpp~is observed. 
\par

\begin{figure}[!h]
    \centering
    \includegraphics[width=0.95\textwidth]{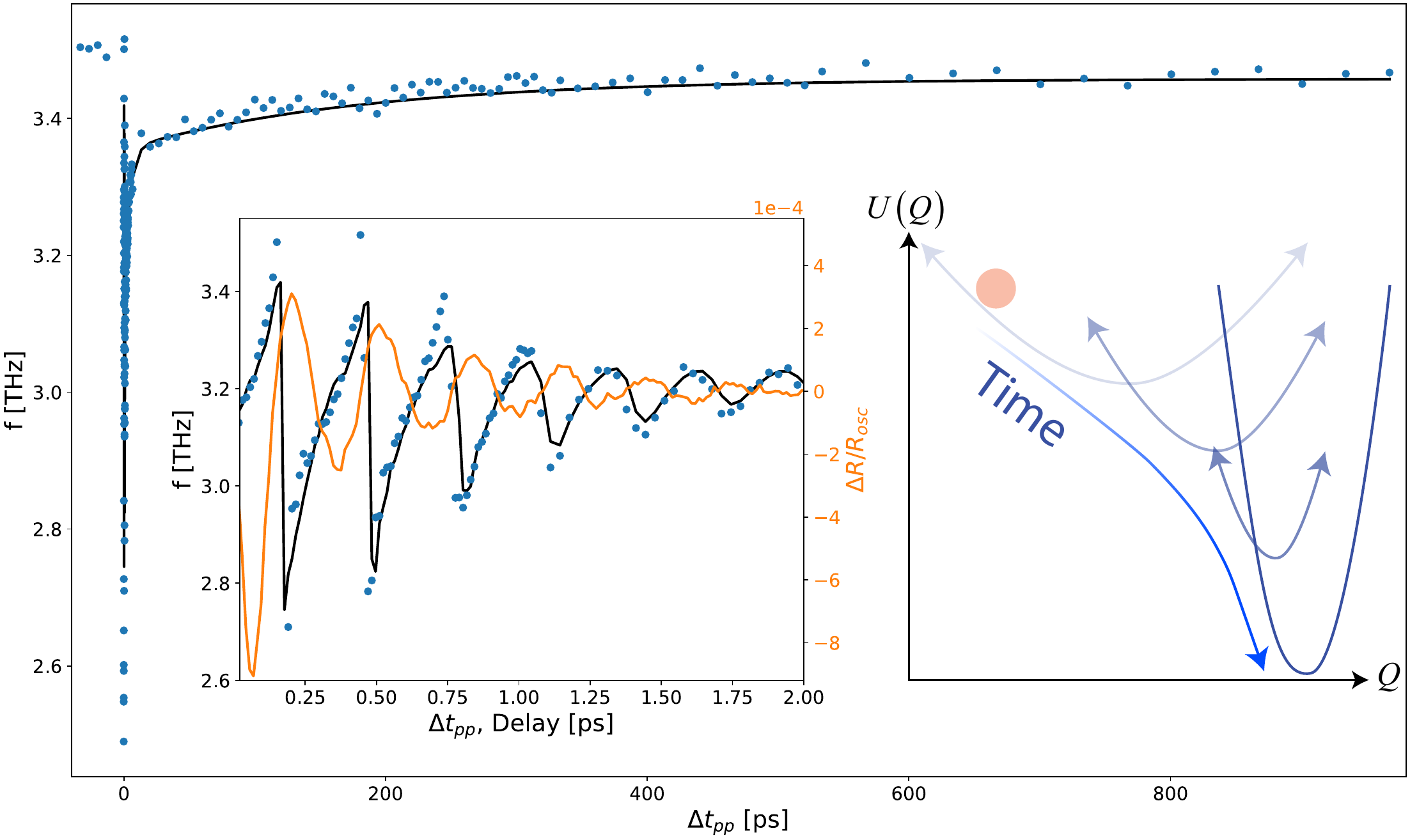}
    \caption{Phonon frequency as a function of \dtpp. Black line is a fit to our oscillator model. In the left inset the blue dots show the data on short timescales for which the horizontal axis represents \dtpp. The orange curve is the oscillatory component of \trR~of the leading pump for which the horizontal axis represents the time delay between the pump and probe pulses, and the black line is a fit to our model. In the inset on the right, an illustration of the DECP process in the presence of electron-phonon coupling. At early times after the photoexcitation, the phonon potential is soft due to the increased electron density (top curve) depicted by the shallow parabola. As time goes by the excited electron density exponentially decays towards its equilibrium value and with it the parabola shifts towards its original position and curvature.}
    \label{Figure 3}
\end{figure}

To closely examine the origin of the nonmonotonic behavior, we plot in Figure \ref{Figure 3} the frequency of the oscillations as a function of \dtpp~(detailed analysis is shown in supplementary section 4). The softening behavior relaxes on longer timescales towards the value obtained when the sample is singly excited by the leading pump.
The (left) inset to Figure \ref{Figure 3} focuses on short timescales of the \dtpp~and shows a distinct periodic non-sinusoidal behavior. The oscillatory component of the \trR~trace obtained when only the leading pump was incident on the sample is plotted for comparison. The periods and time constants of both signals match, indicating that the periodic component of the frequency shift is caused by displacing the atoms along the $A_g$~normal coordinate, serving as direct evidence for anharmonicity. The magnitude of the oscillations reaches values as large as 0.3[THz], which is larger than the magnitude of the observed exponential decay. According to the inverse quadratic behavior predicted for scattering times by Toberer \textit{et al.}, this can change Umklapp scattering times by almost 20\%\cite{toberer2011phonon}. Moreover, this value is larger than what was observed for a similar value of the fluence in the single pump experiment. The periodic signal is asymmetric between its rising and descending parts. The rising side is slow, smooth, and seems to have an inflection point at the times where the \trR~signal crosses zero. In contrast, the descending part is sharp, and its exact features are beyond our experimental resolution, which is limited by the duration of the two pump pulses. The asymmetry between the rising and descending parts of the data indicates that it is not only the position of the atoms along the phonon coordinate that determines the observed frequency but also the direction of their momentum at the time of photo-excitation. The origin of symmetry breaking is the directional nature of the DECP process, i.e., mode extension accompanied by softening and contraction by hardening. Similar asymmetric line-shapes were observed in the phase of coherent phonons in diamond\cite{sasaki2018coherent} and in the phonon lifetime in bismuth\cite{cheng2017coherent} which were also excited using the DECP process. We note that while the similar line-shapes are characteristic of the DECP process, lattice anharmonicity was not observed in either bismuth or diamond. To further investigate the role of the DECP excitation mechanism, we can fit our data to an extension of the DECP model that includes coupling between the resonant frequency and the optical excitation which we observed in SnTe as presented in \ref{Figure 1} (see details in supplementary section 5). This model is depicted in the right inset to Figure \ref{Figure 3}, and describes the dynamics of an oscillator in a parabolic potential which is softened and shifted compared to its equilibrium position. As time goes by, the optically induced transients (increased electron density and elevated temperature) exponentially decay, and the parabolic potential returns to its original curvature and position. The black lines in Figure \ref{Figure 3} and its (left) inset show the fit of our data to this model. The model captures most of the details in our measurements, including the asymmetric lineshapes, the oscillatory behavior of the frequency at short timescales, and the long exponential tail. From the fit, we extract the amplitudes and time constants of the exponential transients, most importantly the electronic component, identified by its short timescale. The electronic timescale is comparable to electronic relaxation timescales measured in the past by time-resolved photo-emission spectroscopy\cite{ito2020observation}. Using the mean-field approximation on the Fr\"{o}lich hamiltonian\cite{frohlich1937theory}, we estimate the electron-phonon coupling constant in SnTe by dividing the electronic frequency shift by the absorbed fluence per unit cell as $\frac{\hbar \Delta\omega}{\Delta n}=0.41[meV]$.
Since the model is fully quadratic and lacks higher order anharmonic terms, one can associate the large anharmonic effects present in our measurements as light-induced anharmonicity caused by the joint action of the DECP mechanism  and the change in the curvature of the parabolic potential for the phonon. Simulations shown in supplementary section 5 show that in the absence of any of these two components, anharmonicity does not emerge. Our model fits the data well both on short and long timescales. However, small discrepancies between the data and fit are consistently observed when the oscillation amplitudes are large. These could originate from anharmonic terms in SnTe not included in our model. We note that while our optical measurements are limited to phonons at the gamma-point, there is no inherent limitation for the mechanism to tune the frequency and anharmonicity of modes away from gamma. Similar measurements were taken using different fluences for the leading and trailing pump pulses. The full datasets are provided in supplementary section 6 and show similar results. The signals vary in their amplitude and characteristic timescales, demonstrating the potential of the double pump method to control material properties.

\begin{figure}[htbp]
    \centering
    \includegraphics[width=0.95\textwidth]{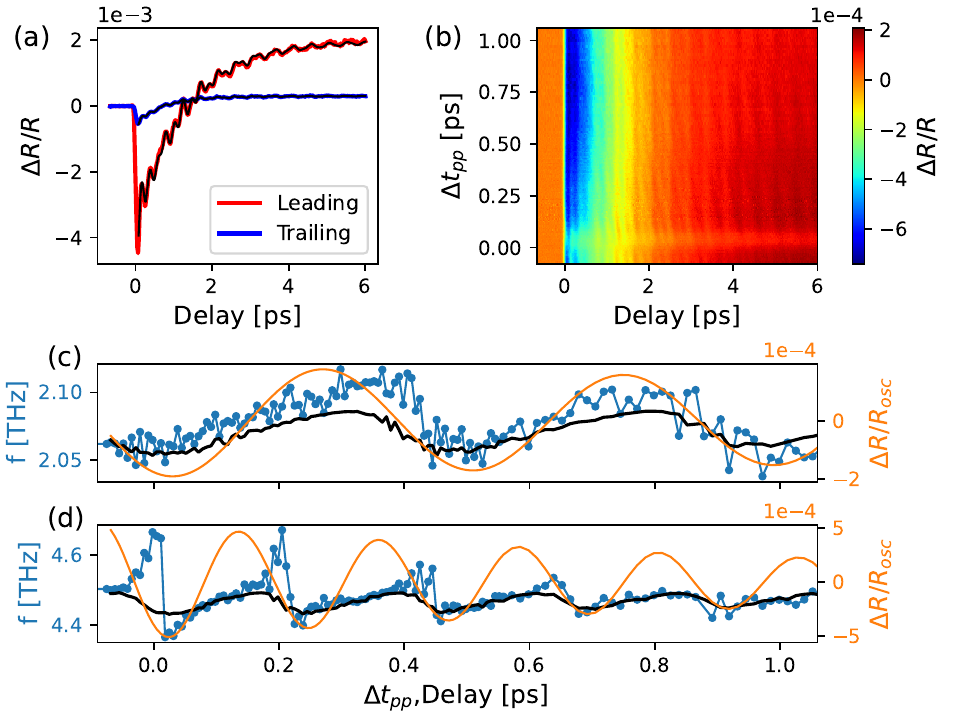}
    \caption{(a) \trR~data in SnSe, taken with single pump excitation. The blue curve is taken only with the trailing pump, whereas the red curve is taken only with the leading pump. (b) Color map of the double pump \trR~data for SnSe. Colors represent the magnitude of \trR, the horizontal axis is the time delay between the probe and the trailing pump, whereas the vertical axis is \dtpp. In panels (c) and (d), similar plots to that presented in Figure \ref{Figure 3} are presented, corresponding to two of the $A_g$ phonons of SnSe. In both panels, the blue curves show the phonons' frequency as a function of \dtpp, the black curves show fit to our model, and the orange curves are the corresponding oscillatory component of \trR~of the leading pump, which were extracted by fitting to the pump-probe data shown in panel (a), as a function of the time delay between the pump and the probe pulses.}
    \label{Figure 4}
\end{figure}

\par
SnTe has the rocksalt structure, therefore its fully symmetric Ag mode is similar to a uniform expansion of the crystal which also happens as a function of increased temperature. Therefore, the observed changes in frequency could be thought of as the coherent and thermal versions of the lattice expansion effect. To test whether the double pump method can identify anharmonicity along less symmetric phonon modes we performed similar measurements on SnSe, a low-symmetry variant of SnTe. Figure \ref{Figure 4}(a) shows the \trR~signal taken with single pump excitation. It is clear from the beat-like pattern, that more than a single frequency is present in the signal corresponding to more than one phonon in agreement with previous measurements of SnSe\cite{huang2022observation}. Figure \ref{Figure 4}(b) shows \trR~as a function of time delay and \dtpp. Unlike in Figure \ref{Figure 2}(b), the presence of multiple frequencies in the data makes it challenging to decipher the anharmonic effects directly from the plot. To decrease the number of fitting parameters, we first subtract the exponential background and fit the data to a combination of damped oscillations to extract the mode frequencies as a function of \dtpp. The results for the two most dominant phonons are shown in Figure \ref{Figure 4} (c-d). Both modes were observed previously in single pump experiments\cite{huang2022observation} and are known to soften with increased fluence. However, as \dtpp~is increased, both modes exhibit periodic modulations at the frequency excited by the leading pump for the same mode. We extract mode-specific electron-phonon coupling for the two phonons in a similar manner as was done in SnTe and obtain $0.10 [meV]$ and $0.15 [meV]$ for the 2.1 THz and 4.5 THz modes, respectively. Similar to SnTe, the model does not capture the large frequency shifts at the peaks of the oscillations, which may indicate contributions from higher-order anharmonic terms. In contrast to SnTe, the lower symmetry of SnSe differentiates between lattice expansion and motion along the crystal's $A_g$ modes, and Huang \textit{et al.}\cite{huang2022observation} even showed that their excitation is entwined with a novel lattice instability towards a higher symmetry phase.

\par
Our experiment serves as a direct measurement of light-induced anharmonicity of a single Raman mode. We present a powerful tool to measure mode-specific electron-phonon coupling. We show that the joint action of the DECP mechanism and electron-phonon coupling leads to effective light-induced anharmonicity. Furthermore, our method could also be used to measure the intrinsic anharmonicity of Raman modes, in contrast to line broadening and coherent coupling between different modes, which are typically used to characterize anharmonicity. Our study demonstrates that double pump-probe spectroscopy is a powerful tool for disentangling the thermal, electronic, and anharmonic contributions to phonon frequency shifts. By applying this technique, we have been able to gain detailed insights into the complex interplay of these factors in SnTe and SnSe. Importantly, the applicability of our approach to disentangling electronic from thermal effects in ultrafast measurements is broader than the application discussed here. For example, light-induced phase transitions are tuned as a function of the pump fluence\cite{de2022decoupling}. In many cases the nature of the transition, whether of thermal or electronic origins is debated. Typically, a theoretical understanding of the particular system at hand is required in addition to the experimental data to reach a definitive conclusion. We show that the double pump technique separates between thermal and electronic effects by their distinct temporal behavior and can greatly aid in solving these debates.

\par

\newpage

\begin{addendum}
\item [Methods] SnTe single crystals were grown using the vapor transport technique without the aid of a transport agent similar to the procedure described in Ref\cite{kannaujiya2020growth}. Stoichiometric amounts of Sn(99.995\% Sigma Aldrich) and Te(99.999\% Thermoscientific) were weighed and thoroughly mixed inside an alumina crucible. The starting materials were sealed under vacuum in a 20 cm long quartz ampule which was placed in a gradient tube furnace. The starting materials were positioned in the hot zone of the furnace. The source and growth zones of the furnace were ramped to 797\textdegree C and 747\textdegree C respectively at a rate of 25\textdegree C/hour. The synthesis took place for 1 week after which large single crystals were found in the cold zone of the tube. A similar procedure was used to grow SnSe. Selected crystals were ground to a fine powder using a mortar and pestle which was used to characterize the structure of the crystals using powder X-ray diffraction (see supplementary section 2). 
\par
Detailed information about the optical experiment, not provided in the main text, is shown in supplementary section 1.

\item A.R. acknowledges support from the Zuckerman Foundation, and the Israel Science Foundation (Grant No. 592/25). This research was supported by the Ministry of Innovation, Science and Technology, Israel. This research was supported by the Ministry of Energy, Israel. We thank the TAU light matter interaction center for supporting this research. Gili Scharf acknowledges support from the Israeli Clore fellowship and Nova Ltd. We thank Shachar Fraenkel, Hadas Soifer, Dominik Maximilian Juraschek and Steven Johnson for fruitful discussions and Yohai Bar Sinai for giving us access to computer resources.

\item [Data availability] The data that support the findings of this study are available from the corresponding author upon reasonable request. 

\item [Author contribution] All authors contributed to the production of this manuscript.
\item[Competing Interests] The authors declare no competing financial interests.
\item[Correspondence] Correspondence and requests for materials
should be addressed to Alon Ron~(email: alonron@tauex.tau.ac.il).
\end{addendum}

\bibliography{Bibliography_SnTe}

\newpage

\setcounter{figure}{0}

\vspace{1cm}
\begin{center}
\begin{spacing}{1.25} 
    \textbf{\large Supplemental Material for: "Coherent control through phonon anharmonicity"}
\end{spacing}
\end{center}

\input{Supplemental/supplementary}

\end{document}

%% file: Supplemental/supplementary.tex


\makeatletter
\let\saved@includegraphics\includegraphics
\renewenvironment{figure}{\@float{figure}}
{\end@float}
\makeatother

\renewcommand{\thefigure}{S\arabic{figure}}





\author{Gili Scharf$^1$, Tomer Hasharoni$^1$, Lara Donval$^1$, Leah Ben Gur$^1$, Alon Ron$^1$}

\maketitle
\begin{affiliations}
 \item Raymond and Beverly Sackler School of Physics and Astronomy, Tel-Aviv 69978, Israel
\end{affiliations}

\section{Optical setup}
This section describes the optical setup used to take the single and double pump probe measurements. 1030 nm pulses were generated by a regenerative amplifier at a repetition rate of 50 kHz, these were used to pump a pair of optical parametric amplifiers seeded by the same white light crystal. The wavelength of the signal beams of the two optical parametric amplifiers were set to 750 nm and 800 nm and were used as pump and probe respectively. For the single pump experiments, the fluence of the probe did not change throughout the experiment and was set to $20 [\mu J/cm^2]$, that of the pump beams was varied as described in the main text. The pump beam was split by a 50:50 beamsplitter. The pump beam was modulated at a frequency of 5 kHz in one of two possible positions before and after the beamsplitter for isolation of the signal originating from the trailing pump beam as described in the main text. The pumps' intensities were varied by using neutral density filters. 

\begin{figure}[!ht]

    \centering
    \includegraphics[width=1\textwidth]{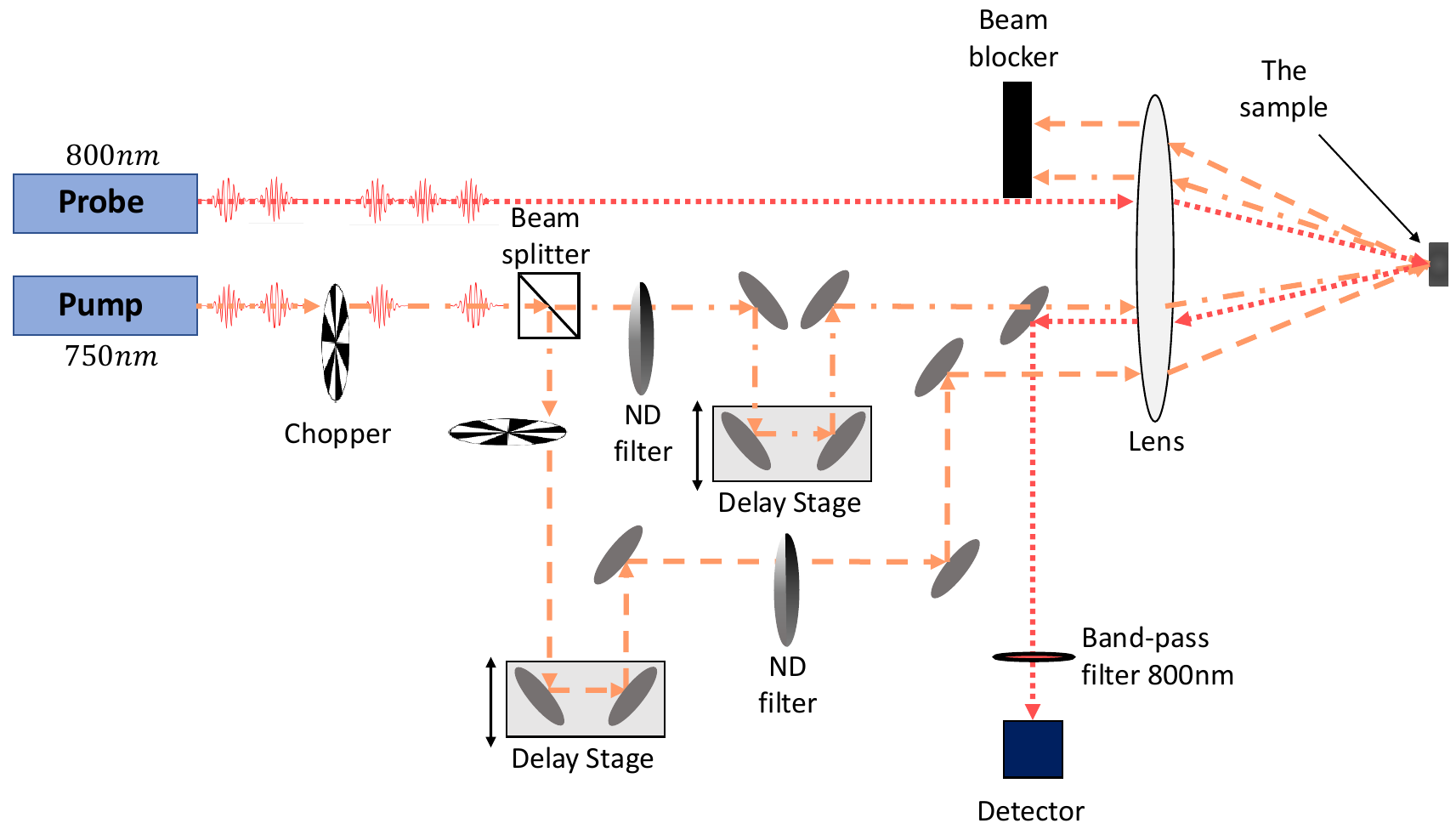}
    \caption{Schematic of the optical setup used in this work. Unlabeled components are silver mirrors.}
    \label{DPP setup figure}
\end{figure}

\clearpage

\section{X-ray diffraction}
\begin{figure}[!ht]

    \centering
    \includegraphics[width=1\textwidth]{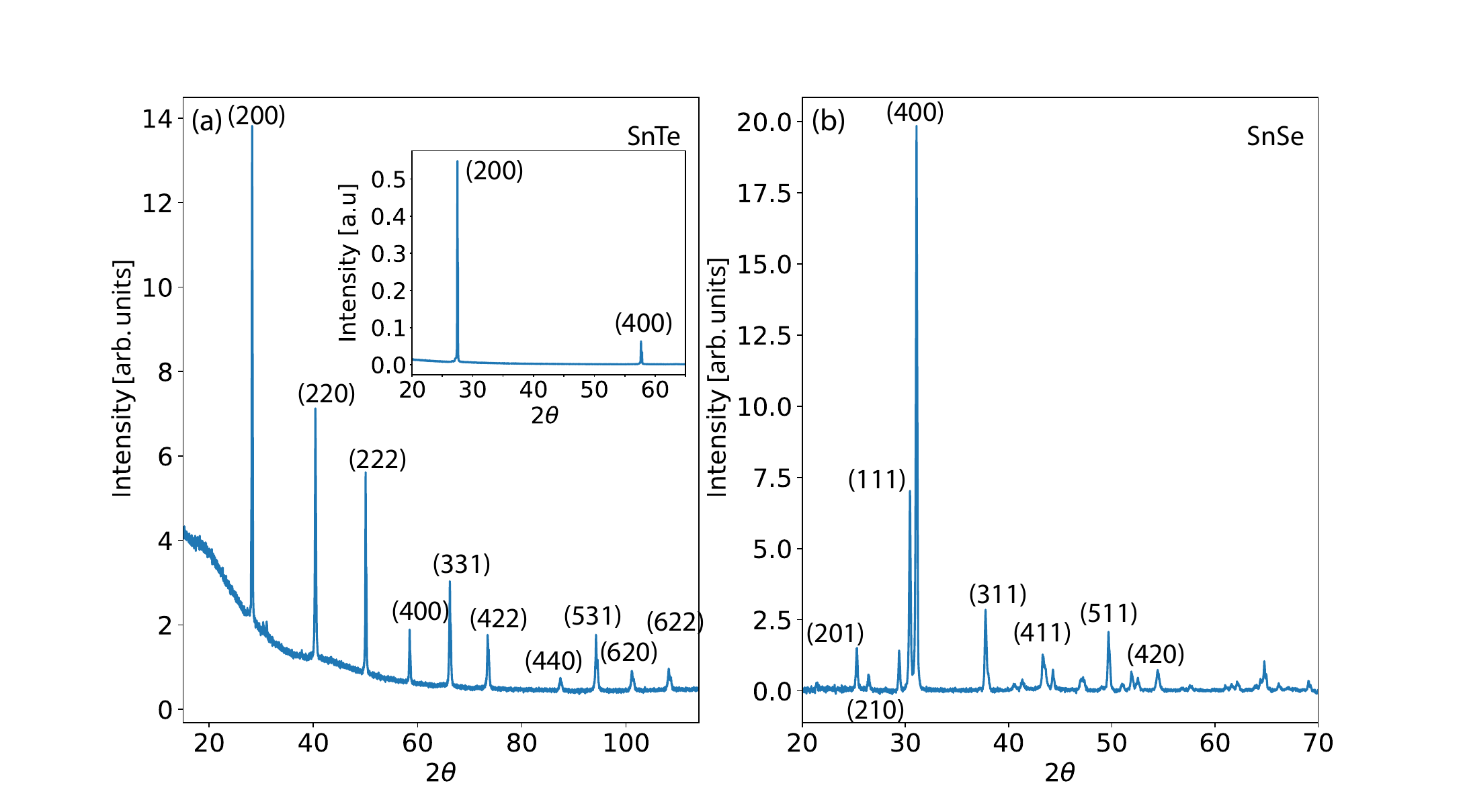}
    \caption{Powder X-ray diffraction data of (a) SnTe and (b) SnSe samples are presented. The inset of panel (a) shows the X-ray diffraction from the SnTe crystal that was used in the measurements presented in the main text. Miller indices are displayed next to the main observed peaks. For SnSe, a polynomial background was subtracted from the data.}
    \label{XRD figure}
\end{figure}

In this section, we show the X-ray diffraction data of SnTe and SnSe powders taken from the same batches as the crystals used for the optical experiments. The crystals were ground into powder using an agate mortar and pestle before the measurements. All observable peaks were associated with peaks expected for the known phases of SnTe and SnSe using Rietveld refinement. Figure \ref{XRD figure} shows the measured diffraction patterns taken using a Cu k $\alpha$~source. All peaks in the pattern agree with the expected peak positions of the known phases, and weak peaks are not marked for clarity. A polynomial background was subtracted from the SnSe data. The inset to panel (a) shows a measurement on the SnTe crystal used in the measurements presented in the main text. The presence of (200) and (400) peaks only, indicates that the surface is the (100)surface.

\clearpage

\section{Estimation of the thermal time constant}\label{Estimation of the thermal time constant section}

We estimate the thermal time constant by:
\begin{equation}
\tau_{thermal}  = R_{thermal}  \cdot C_{thermal}
\label{tau_thermal eq}
\end{equation}
Where $R_{thermal}$ and $C_{thermal}$ are the thermal resistance and heat capacity of the probed volume, respectively. We can express the $R_{thermal}$ in the following manner:
\begin{equation}
\rho_{thermal} = 1/\kappa = R_{thermal} \frac{A}{l} \Rightarrow R_{thermal} = \frac{l}{A \cdot \kappa}
\label{R_thermal eq}
\end{equation}
Where $\rho_{thermal}$, the thermal resistivity, is the inverse of $\kappa$, the thermal conductivity, $A$ is the surface area of the incident probe beam and $l$ is the penetration depth of the probe. The heat capacity can be expressed as the heat capacity per unit mass, $c_{thermal}$, multiplied by $M$, the mass of the volume irradiated by the probe beam:
\begin{equation}
C_{thermal} = c_{thermal} \cdot M = c_{thermal} \cdot \rho \cdot V = c_{thermal} \cdot \rho \cdot A \cdot l
\label{C_thermal eq}
\end{equation}
The penetration depth of the probe, which is the inverse of the attenuation constant ($\alpha$), is determined by the wavelength of the probe ($\lambda$) and the imaginary part of the complex refractive index ($\tilde{n}(\lambda)$):
\begin{equation}
l = \frac{1}{\alpha} =  (\frac{4\pi}{\lambda} \cdot \Im[\tilde{n}(\lambda)])^{-1}
\label{penetration depth eq}
\end{equation}

Combining all of the above and using values from literature\cite{Suzuki_1995, beattie1969temperature, C4CP02091J, damon1966thermal} gives $\tau_{thermal} \sim$ 25.02 ps.

\clearpage

\section{Detailed analysis}\label{analysis of the DPP dada section}

Analysis of the data shown in Figure 3 in the main text was performed in the following way. A high pass filter with a cutoff frequency of 2 THz was first applied to the data. The oscillatory component was fit according to the following expression:

\begin{equation}
Ae^{-t/\tau}cos(\omega t+\phi)+c.\label{eqS1}
\end{equation}

\begin{figure}[htbp]
    \includegraphics[width=1\textwidth]{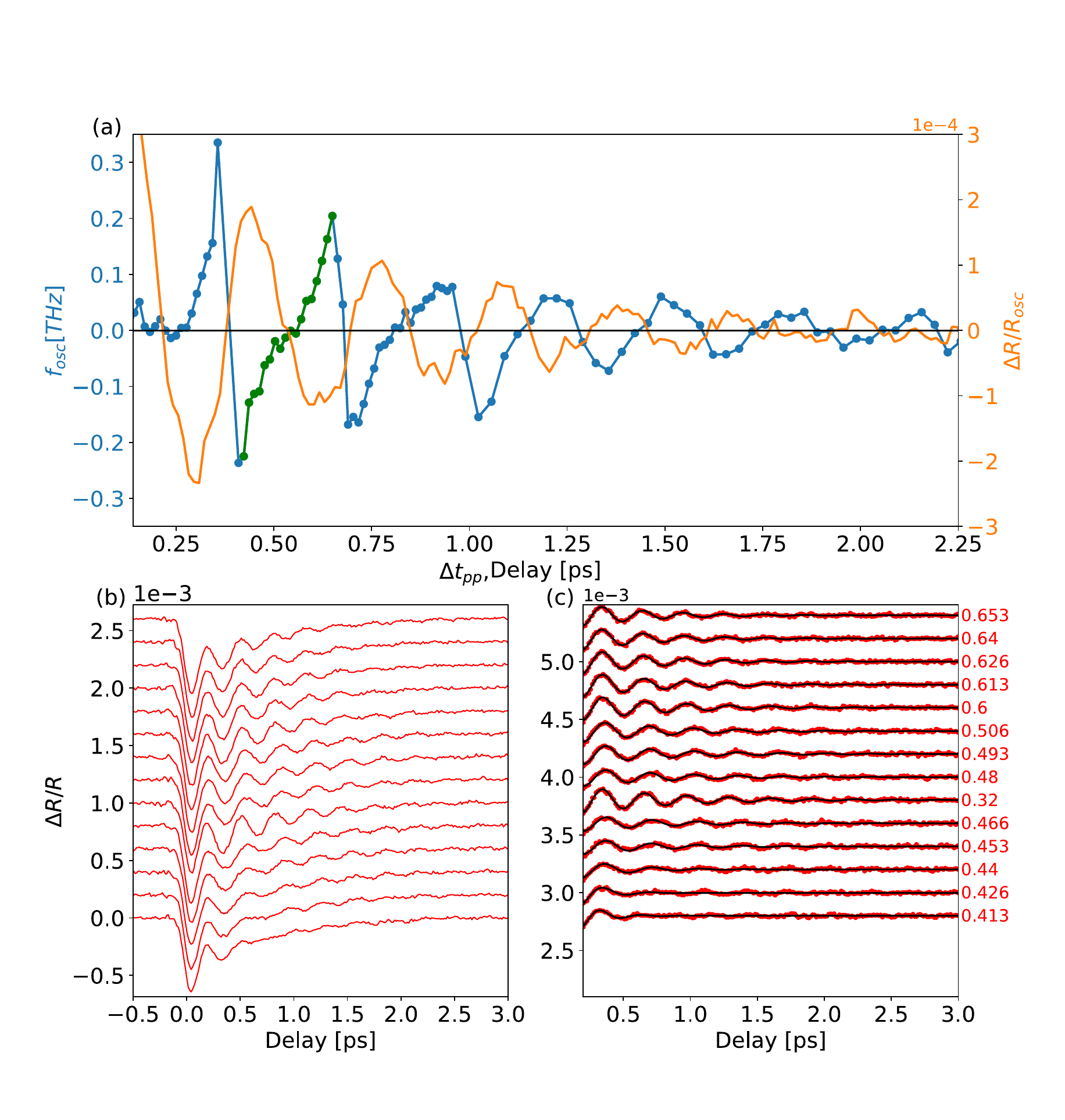}
    \vspace{-55pt}
    \caption{(a) Same data as in Figure 3 in the main text. A detailed analysis for the points highlighted in green is presented in panels (b, c). (b) Raw data vertically shifted for clarity. Values of \dtpp~are the same as the corresponding traces in panel (c). (c) The data in (b) after the application of a high-pass filter with a cutoff frequency at 2 THz. The numbers to the right of the panel correspond to the values of \dtpp~in ps. Black lines are the fit results to supplementary equation (\ref{eqS1}).}
    \label{Figure S3}
\end{figure}

The final results of this analysis procedure are shown in Figure 3 of the main text. In the following paragraph, we provide a detailed analysis on a portion of the data. Figure \ref{Figure S3}(a) shows the same result as plotted in the main text Figure 3 with one distinction. A subset of the data is highlighted in green and its detailed analysis is presented in this section. Figure \ref{Figure S3}(b) shows the raw data shifted for clarity for the range \dtpp~between 0.4 to 0.65[ps]. The red dots in Figure \ref{Figure S3}(c) show the same data after the application of a high-pass filter with a cutoff frequency of 2 THz. The numbers to the right of panel (c) are values of \dtpp~corresponding to the data near them in ps. The black lines show fits to the data according to supplementary equation \ref{eqS1}. As can be seen, the expression fits the data nicely for all values of \dtpp.

\clearpage

\section{Modeling the DECP excitation}

\begin{figure}[htbp]
    \includegraphics[width=1\textwidth]{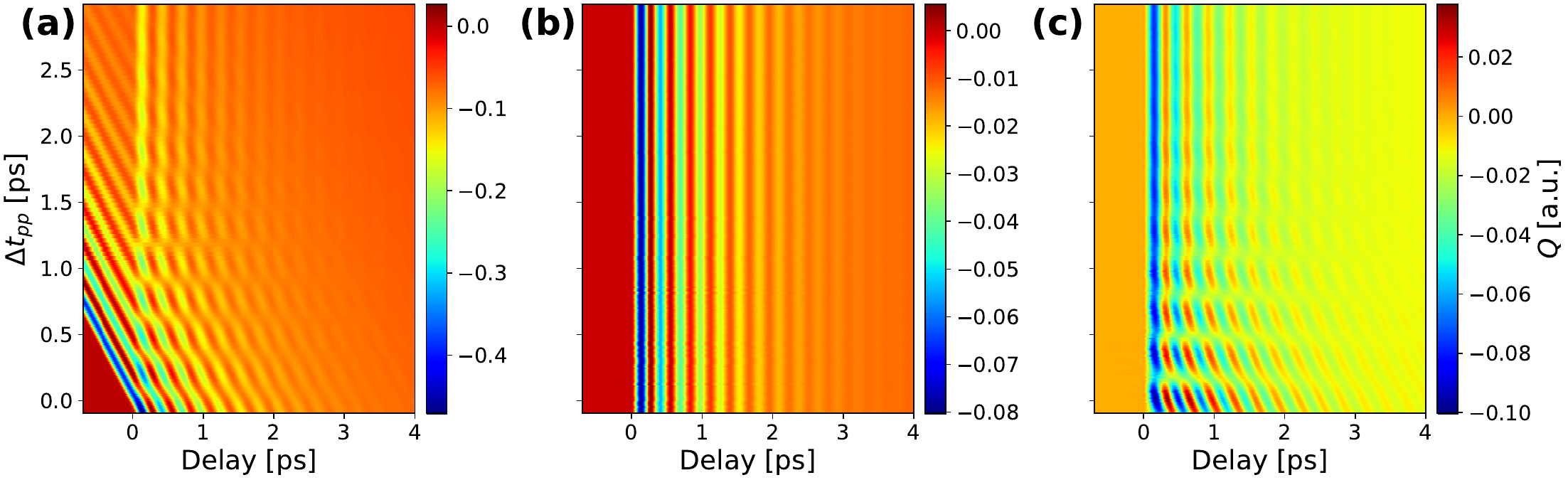}
    \vspace{-45pt}
    \caption{Simulation of a double pump experiment according to the model (a) in the case $\Delta\omega=0$~without a modulation on the trailing pump, (b) with modulation on the trailing pump and (c) with modulation on the trailing pump and $\Delta\omega\neq 0$.}
    \label{DPP simulations}
    \label{DPP simulation with df=0}
\end{figure}

In the DECP mechanism, the excitation of the phonon is induced by a change in the equilibrium position, or the minima of the parabolic potential for the phonons along the phonon coordinate $Q$. The equation of motion for $Q$, as a response to a pump excitation  $f_{exc}$ is:
\begin{equation} \label{regular DECP equation of motion}
    \ddot Q = -\omega_0^2(Q-\Delta Q f_{exc}) - \frac{1}{\tau_p} \dot Q  
\end{equation}
Where  $\Delta Q$ is a proportionality constant between the shift in equilibrium position and the instantaneous intensity of the pump excitation,  $\tau_p$ is the phonon's lifetime and $\omega_0$~is its resonant frequency. The dynamics produced by this equation are fully harmonic and result in a typical cosine-like behavior as evident in Figure \ref{DPP simulation with df=0}. 

Assuming that the excitation induced by the pump is instantaneous, we can express $f_{exc}(t)$ as a step function which multiplies a sum of decaying exponents. Each exponent represents a different contribution to the phonon excitation (e.g., electronic and/or thermal contributions): 
\begin{equation}\label{f_exc general eq}
    f_{exc}(t)=\Theta(t-t_0)\sum_i A_ie^{t-t_0/\tau_i}
\end{equation}
 In SnTe, our single pump experiment shows that the excitation consists of three different contributions with three different time-scales (See Figure 1 in the main text). Thus, in our model, we sum over three contributions arising from the excitation of electrons, thermal interband electronic relaxation and long timescale thermal relaxation of the sample.
In general, the phonon frequency $\omega_0$~can depend on the instantaneous temperature, carrier density, etc. This results in the following equation of motion:

\begin{equation}\label{single phonon equation of motion}
    \ddot Q = -(\omega_0 - \Delta\omega f_{exc})^2(Q-\Delta Q f_{exc}) - \frac{1}{\tau_p} \dot Q  
\end{equation}
where $\Delta\omega$ represents the coupling between the phonon frequency and the pump excitation.

These equations of motion could be numerically solved for a given set of parameters to simulate the motion of an oscillator under a DECP excitation similar to that in our experiment. In a double pump experiment, the double excitation can is modeled as the sum of two single pump excitations shifted by the variable time delay \dtpp:
\begin{equation}\label{f_exc DPP eq}
    f_{exc}(t)=f_{leading}(t+\Delta t_{pp})+f_{trailing}(t)
\end{equation}

\begin{figure}[htbp]
    \includegraphics[width=1\textwidth]{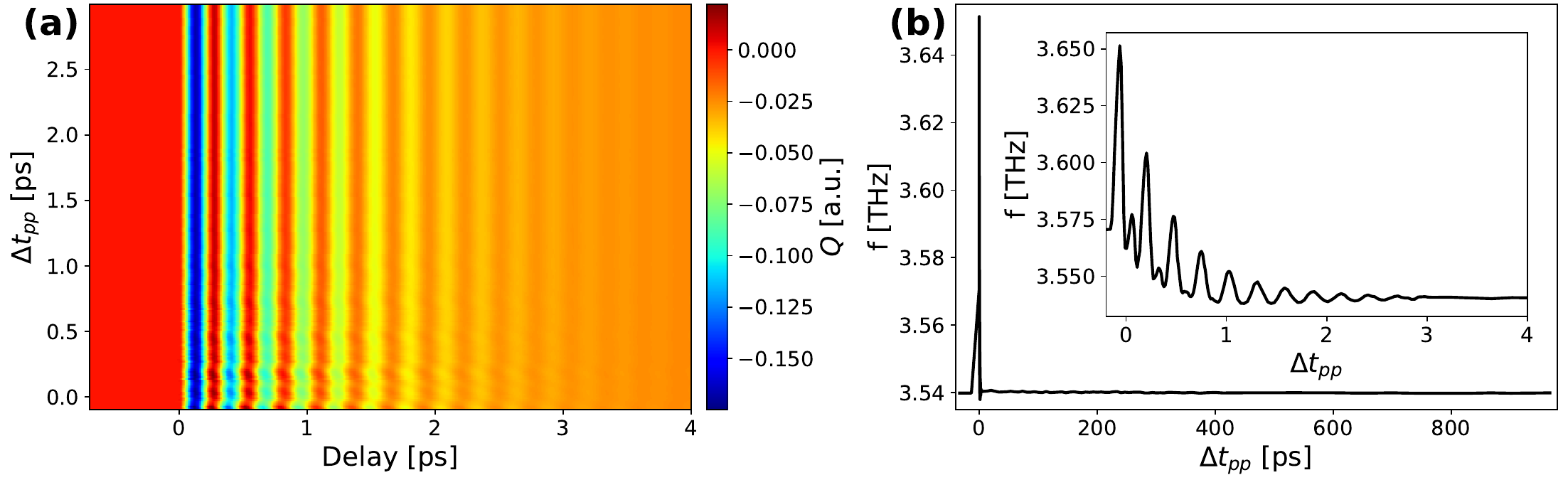}
    \vspace{-45pt}
\caption{Double pump simulation with a cubic anharmonic term in the equation of motion. (a) The colors show the resultant displacement from the equilibrium position along the phonon coordinate $\Delta Q$~with modulation of the trailing pump. (b) The extracted frequency as a function of \dtpp~for the simulation presented in panel (a).}
    \label{DPP simulations with anharmonicity}
\end{figure}

To fit our experimental results to the model, we mimic the modulation of the trailing pump by the mechanical chopper by subtracting the results calculated with and without the trailing excitation. The results of this fit are shown in Figure 3 and 4 of the main text. To show the generality of our method in detecting any form of anharmonicity, we show in Figure \ref{DPP simulations with anharmonicity} a simulation which includes a cubic anharmonic term in the equation of motion. To emphasise the importance of using the chopper to modulate the trailing pump, we show in Figure \ref{DPP simulations} the results of simulating a double pump experiment with and without modulation of the second pump with $\Delta\omega=0$~(panels a and b)and $\Delta\omega\neq0$ (panel c). For $\Delta\omega=0$~(panel a), the simulation shows that modulation of the oscillatory peaks as \dtpp~is scanned can originate by the interference of the two oscillatory components excited by the two pump pulses (panel a), which could be easily mistaken for anharmonicity. In panel b, we show results calculated with the modulation of the trailing pump and $\Delta\omega=0$. The oscillatory peaks appear at constant values of \dtpp, indicating the absence of any form of anharmonicity as expected for the case of $\Delta\omega=0$. In contrast, when $\Delta\omega\neq0$ (panel c), peak modulations are observed in the trailing pump modulated case and light-induced anharmonicity is verified.

\clearpage

\section{Double pump measurements of SnTe with additional fluences}

\begin{figure}[htbp]
    \includegraphics[width=1\textwidth]{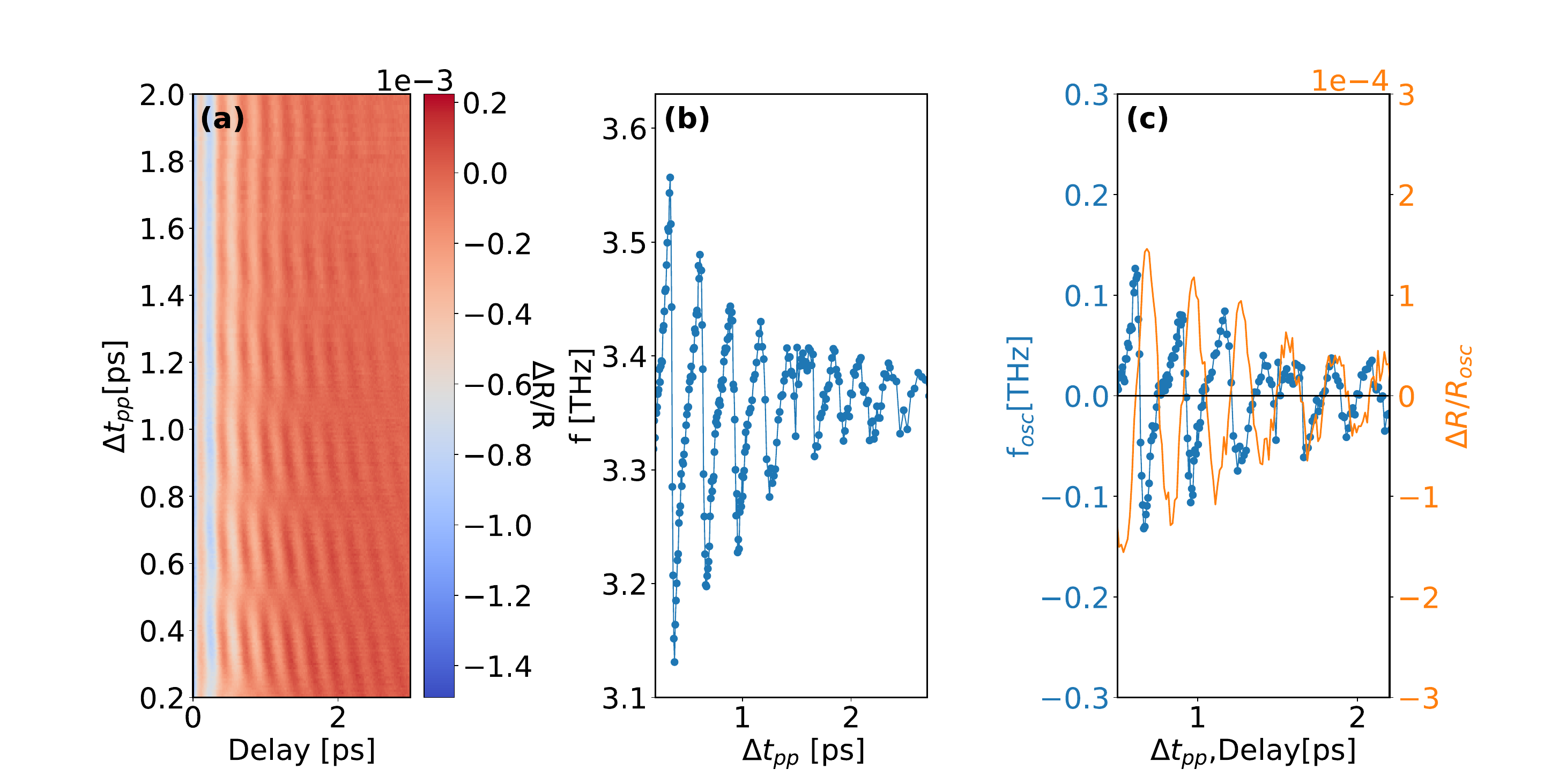}
    \caption{(a) Colormap showing \trR~as a function of delay time and \dtpp~for incident fluences of 1.454 and 0.708$[mJ/cm^2]$. (b) The extracted phonon frequencies as a function of \dtpp. (c) The blue curve shows the oscillatory component of the measured frequencies as a function of \dtpp. The overlaid orange curve shows the oscillatory component of \trR~measured using only the leading pump.}
    \label{DPP run 4 figure}
\end{figure}

\begin{figure}[htbp]
    \includegraphics[width=1\textwidth]{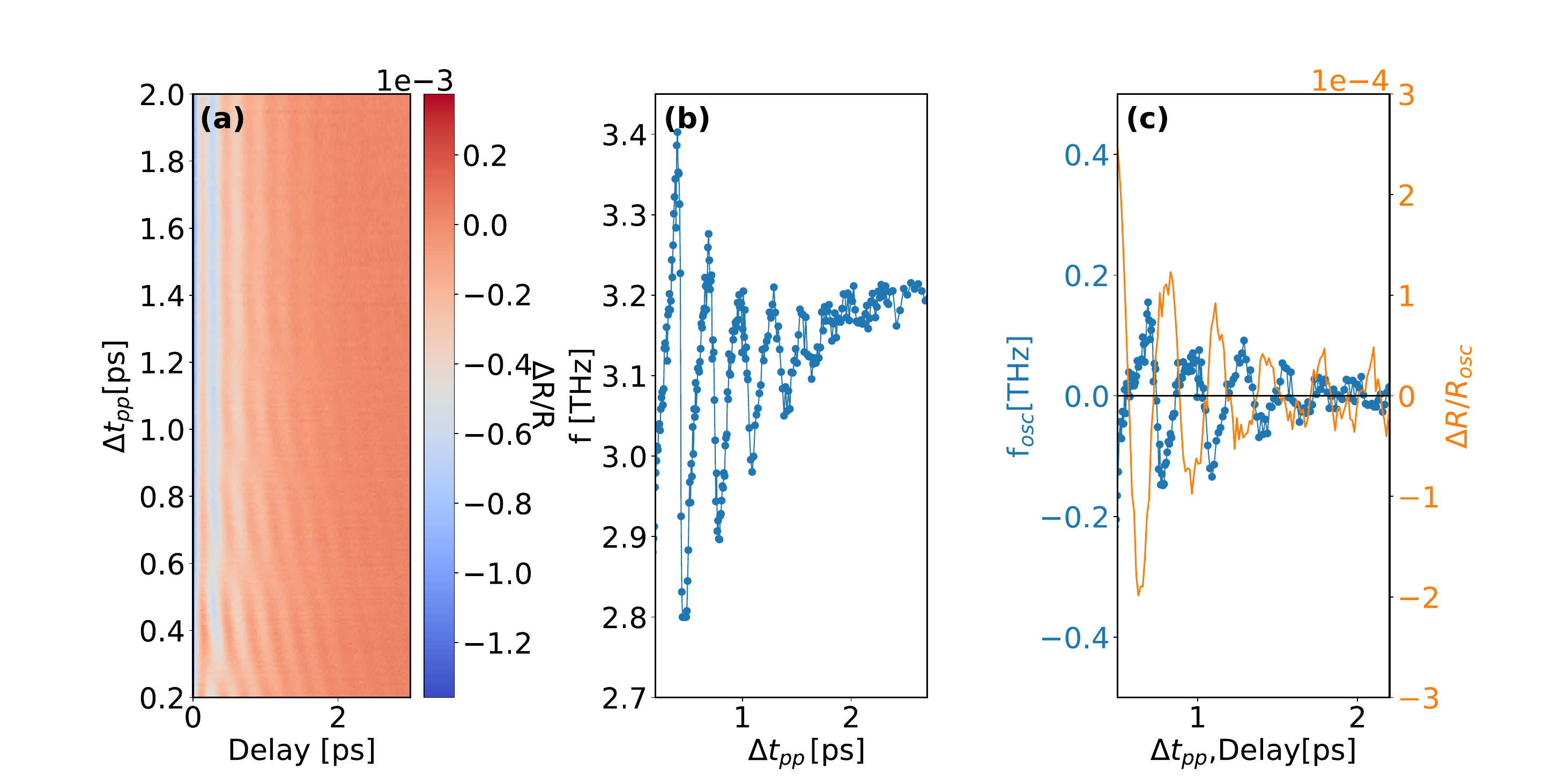}
    \caption{(a) Colormap showing \trR~as a function of delay time and \dtpp~for incident fluences of 3.51 and 0.777$[mJ/cm^2]$. (b) The extracted phonon frequencies as a function of \dtpp. (c) The blue curve shows the oscillatory component of the measured frequencies as a function of \dtpp. The overlaid orange curve shows the oscillatory component of \trR~measured using only the leading pump.}
    \label{DPP run 3 figure}
\end{figure}

The main text presents double pump measurements of SnTe at a single combination of fluences. In this section, we show two additional datasets, the first with incident fluences of 1.454 and 0.708$[mJ/cm^2]$ (corresponding to excitation densities of $1.509\times10^{21}$ and $7.35\times10^{20} [cm^{-3}]$\cite{Suzuki_1995}) for the leading and trailing pump beams respectively, and the second with incident fluences of 3.51 and 0.777$[mJ/cm^2]$ (which correspond to excitation densities of $3.639\times10^{21}$ and $8.06\times10^{20} [cm^{-3}]$\cite{Suzuki_1995}). The measurements show qualitatively similar results to those shown in the main text. The main difference is that the frequency of modulations shown in Figures \ref{DPP run 4 figure}(c) and Figure \ref{DPP run 3 figure}(c) are different than the frequency presented in the main text. In all cases it corresponds to the frequency of the phonon observed when using only the leading pump. 

\clearpage

\section{Simulation of chirped phonon and comparison to a non-chirped phonon for SnTe}

\begin{figure}[htbp]
    \includegraphics[width=1\textwidth]{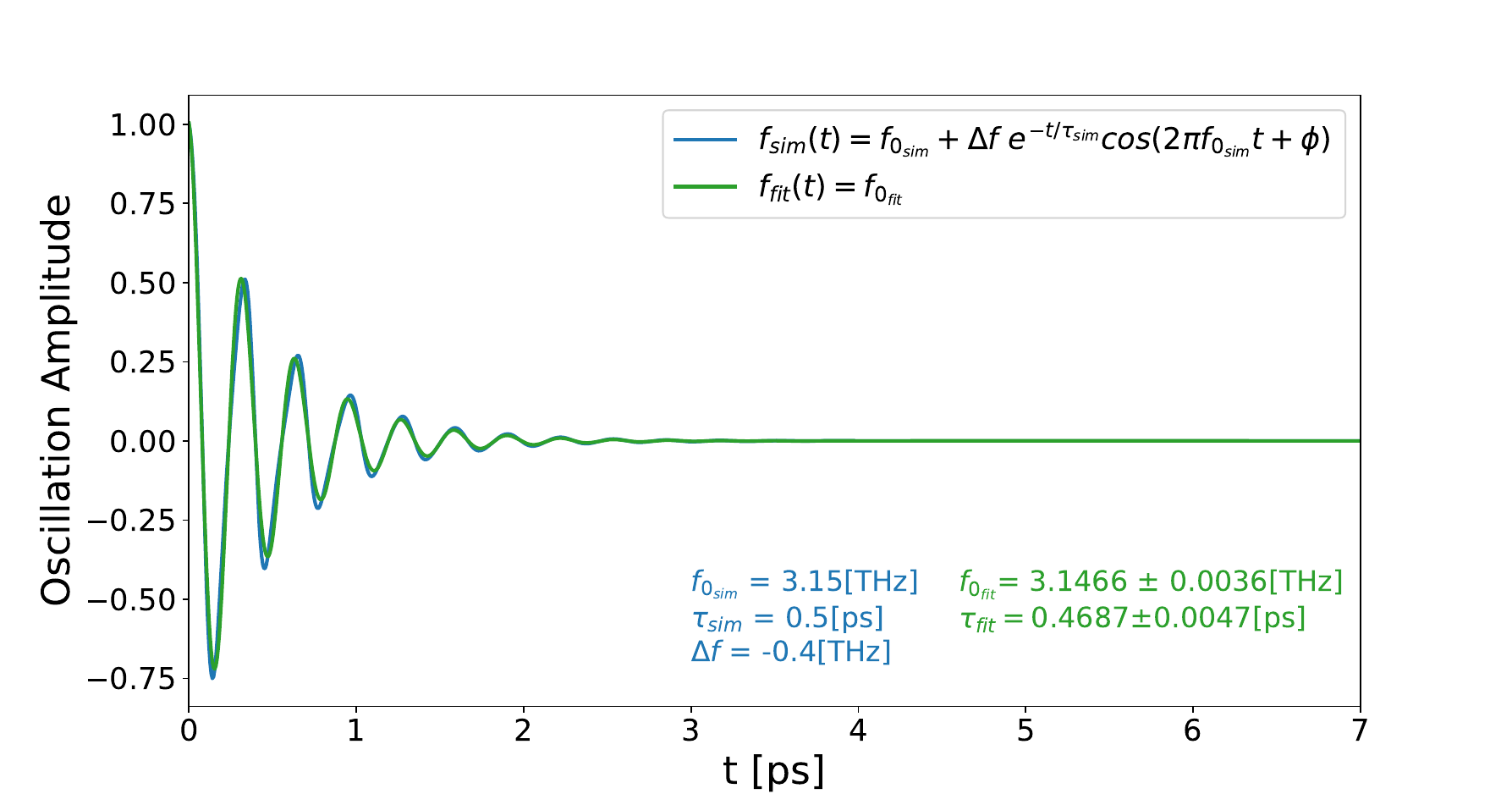}
    \vspace{-45pt}
    \caption{A simulation of a phonon decay with a frequency time-dependency of $f(t)=f_0 + \Delta f\ e^{-t/\tau}cos(2\pi f_0t + \phi)$  is shown in blue. For the simulation, $f_0$ and $\tau$ were selected to be the values from the fit to the pump-probe data of the leading pump, shown in Figure 2(a) in the main text, $\phi$ was taken to be 0, and $\Delta f$ was determined using Figure 3 in the main text. All these values ($\phi$ was set to 0), are shown in the figure in blue. In green, we show a fit to the simulated data without accounting for any frequency modulation. The resulting frequency and time-decay are shown in the figure in green.}
    \label{simulation based on run 2 figure}
    
\end{figure}

\begin{figure}
    \includegraphics[width=1\textwidth]{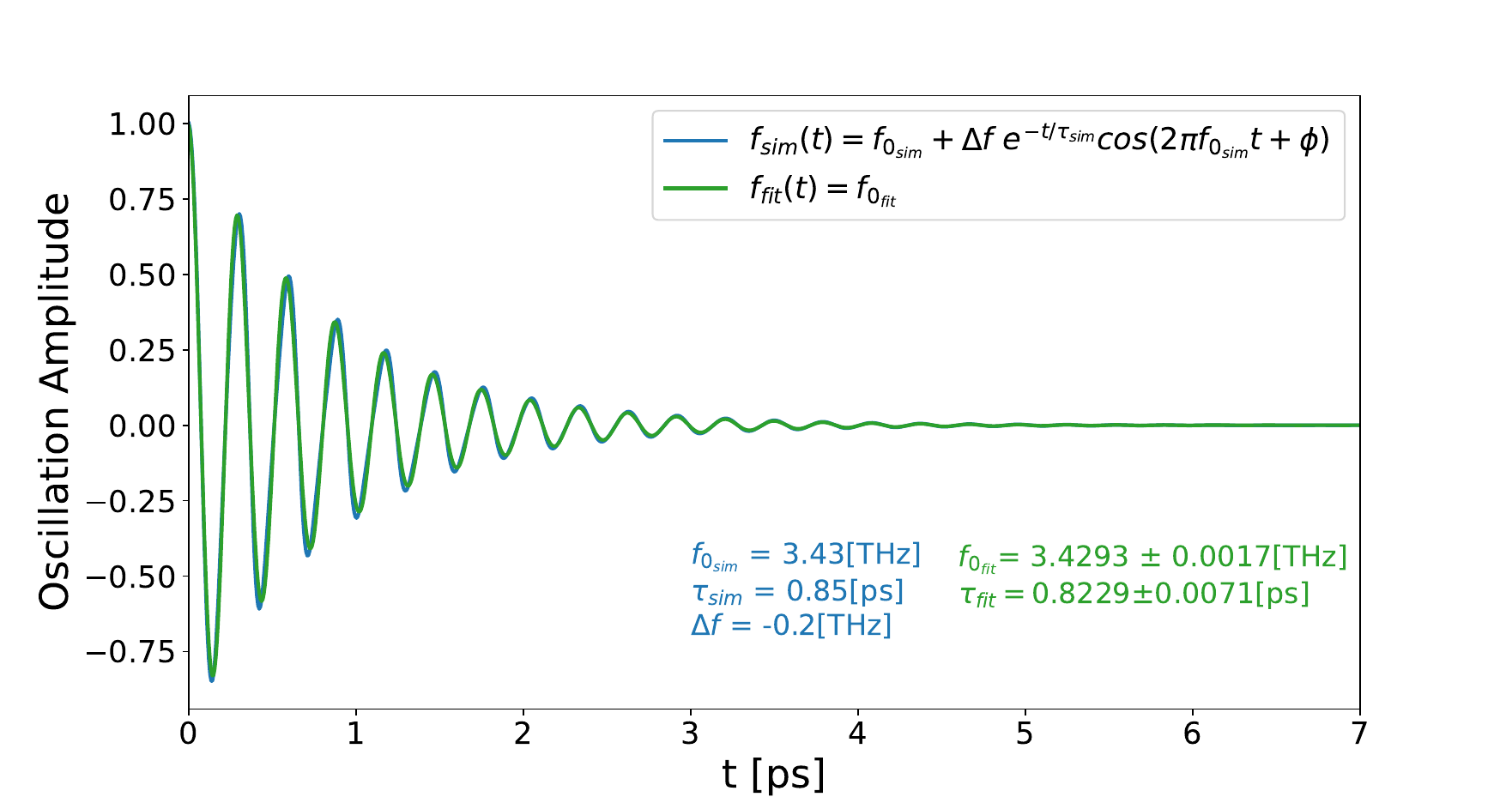}
    \vspace{-45pt}
    \caption{A simulation of a phonon decay with a frequency time-dependency of $f(t)=f_0 + \Delta f\ e^{-t/\tau}cos(2\pi f_0t + \phi)$  is shown in blue. For the simulation, $f_0$ and $\tau$ were taken as the experimental values from the pump-probe data with leading pump fluence of 1.454$[mJ/cm^2]$, $\phi$ was set to zero, and $\Delta f$ was determined using Figure \ref{DPP run 4 figure}(c) in the main text. All these values ($\phi$ was set to 0) are shown in the figure in blue. In green, we show a fit of the simulated data to non-modulated decaying oscillations. The resulting frequency and time-decay are shown in the figure in green.}
    \label{simulation based on run 4 figure}
\end{figure}

In this section, we show that frequency modulations in the phonon frequency of SnTe are undetectable due to the short lifetime of the phonon and that if present, changes to the resulting values will be within the error of the central frequencies reported. We simulate exponentially decaying oscillatory time series with central frequencies similar to those measured in our experiment. We add frequency modulation according to the equation shown in the figure legend. The values for the magnitude of the modulation are taken from the experimental results shown in the main text Figure 3 and in the supplementary, Figure \ref{DPP run 4 figure}(c). The similarity of the modulated signal to the non-modulated signal is clear from the good agreement between them as seen in Figures \ref{simulation based on run 2 figure} and \ref{simulation based on run 4 figure}. Any visible discrepancies are comparable to the magnitude of our experimental measurement error. This means that under these conditions, modulations to the central frequency (such as chirp) are undetectable in our experiment and, if present in SnTe.

\clearpage

\section{Double pump measurements of SnSe with additional fluences}

\begin{figure}
    \includegraphics[width=1\textwidth]{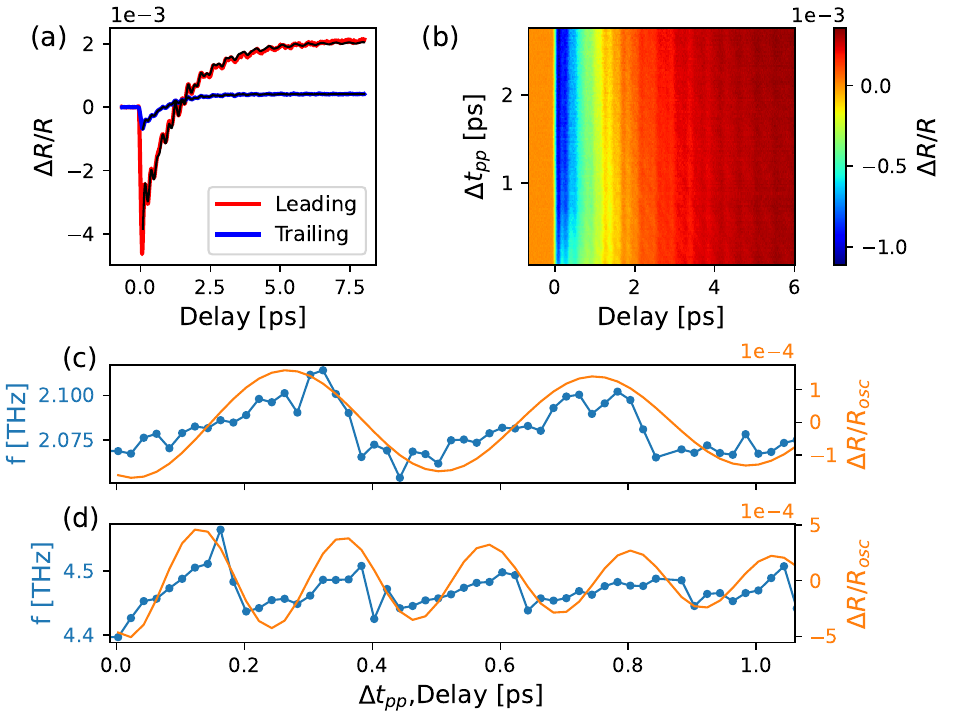}
    \vspace{-45pt}
   \caption{Double pump measurements of SnSe. (a) \trR~data in SnSe, taken with single pump excitation. The blue curve is taken only with the trailing pump whereas the red curve is taken only with the leading pump. (b) Color map of the double pump \trR~data for SnSe. Colors represent the magnitude of \trR, the horizontal axis is the time delay between the probe and the trailing pump whereas the vertical axis is \dtpp. In panels (c) and (d), plots similar to those presented in Figure 3 are presented, corresponding to two of the $A_g$ phonons of SnSe. In both panels, the blue curves show the phonons' frequency as a function of \dtpp~and the orange curves are the corresponding oscillatory component of \trR~of the leading pump, which were extracted by fitting to the pump-probe data shown in panel (a), as a function of the time delay between the pump and the probe pulses.}
    \label{1st DPP run results for SnSe}
\end{figure}

The main text presents double pump measurements of SnSe at a combination of fluences - 3.55 and 1.20$[mJ/cm^2]$ (corresponding to excitation densities of $\approx 1.01\times10^{21}$ and $3.43\times10^{20} [cm^{-3}]$)\cite{barrios2014chemically} for the leading and trailing pumps respectively. In this section, we show an additional dataset with incident fluences of 4.00 and 0.95$[mJ/cm^2]$ (corresponding to excitation densities of $\approx 1.14\times10^{21}$ and $2.72\times10^{20} [cm^{-3}]$)\cite{barrios2014chemically} for the leading and trailing pump beams respectively. The measurements show qualitatively similar results to those shown in Figure 4 in the main text. The main difference is that the frequencies of the modulations shown in Figure \ref{1st DPP run results for SnSe} panels (c) and (d) are different than the frequencies presented in Figure 4 panels (c) and (d) in the main text. In all cases they correspond to the frequency of the phonon observed when using only the leading pump.

\clearpage



%% file: Bibliography_SnTe.bib
@article{teitelbaum2018direct,
  title={Direct measurement of anharmonic decay channels of a coherent phonon},
  author={Teitelbaum, Samuel W and Henighan, Thomas and Huang, Yijing and Liu, Hanzhe and Jiang, Mason P and Zhu, Diling and Chollet, Matthieu and Sato, Takahiro and Murray, {\'E}amonn D and Fahy, Stephen and others},
  journal={Physical review letters},
  volume={121},
  number={12},
  pages={125901},
  year={2018},
  publisher={APS}
}

@article{forst2011nonlinear,
  title={Nonlinear phononics as an ultrafast route to lattice control},
  author={F{\"o}rst, Michael and Manzoni, Cristian and Kaiser, Stefan and Tomioka, Yasuhide and Tokura, Yoshinori and Merlin, Roberto and Cavalleri, Andrea},
  journal={Nature Physics},
  volume={7},
  number={11},
  pages={854--856},
  year={2011},
  publisher={Nature Publishing Group UK London}
}

@article{li2014phonon,
  title={Phonon self-energy and origin of anomalous neutron scattering spectra in SnTe and PbTe thermoelectrics},
  author={Li, CW and Hellman, Olle and Ma, Jie and May, AF and Cao, HB and Chen, X and Christianson, AD and Ehlers, G and Singh, DJ and Sales, BC and others},
  journal={Physical review letters},
  volume={112},
  number={17},
  pages={175501},
  year={2014},
  publisher={APS}
}

@article{wei2021phonon,
  title={Phonon anharmonicity: a pertinent review of recent progress and perspective},
  author={Wei, Bin and Sun, Qiyang and Li, Chen and Hong, Jiawang},
  journal={Science China Physics, Mechanics \& Astronomy},
  volume={64},
  number={11},
  pages={117001},
  year={2021},
  publisher={Springer}
}

@article{lee2014resonant,
  title={Resonant bonding leads to low lattice thermal conductivity},
  author={Lee, Sangyeop and Esfarjani, Keivan and Luo, Tengfei and Zhou, Jiawei and Tian, Zhiting and Chen, Gang},
  journal={Nature communications},
  volume={5},
  number={1},
  pages={3525},
  year={2014},
  publisher={Nature Publishing Group UK London}
}

@article{ribeiro2018strong,
  title={Strong anharmonicity in the phonon spectra of PbTe and SnTe from first principles},
  author={Ribeiro, Guilherme AS and Paulatto, Lorenzo and Bianco, Raffaello and Errea, Ion and Mauri, Francesco and Calandra, Matteo},
  journal={Physical Review B},
  volume={97},
  number={1},
  pages={014306},
  year={2018},
  publisher={APS}
}

@article{golonzka2001coupling,
  title={Coupling and orientation between anharmonic vibrations characterized with two-dimensional infrared vibrational echo spectroscopy},
  author={Golonzka, O and Khalil, M and Demird{\"o}ven, N and Tokmakoff, A},
  journal={The Journal of Chemical Physics},
  volume={115},
  number={23},
  pages={10814--10828},
  year={2001},
  publisher={AIP Publishing}
}

@article{fulmer2004pulse,
  title={A pulse sequence for directly measuring the anharmonicities of coupled vibrations: Two-quantum two-dimensional infrared spectroscopy},
  author={Fulmer, Eric C and Mukherjee, Prabuddha and Krummel, Amber T and Zanni, Martin T},
  journal={The Journal of chemical physics},
  volume={120},
  number={17},
  pages={8067--8078},
  year={2004},
  publisher={American Institute of Physics}
}

@article{golonzka2001vibrational,
  title={Vibrational anharmonicities revealed by coherent two-dimensional infrared spectroscopy},
  author={Golonzka, O and Khalil, M and Demird{\"o}ven, N and Tokmakoff, A},
  journal={Physical review letters},
  volume={86},
  number={10},
  pages={2154},
  year={2001},
  publisher={APS}
}

@article{lin2022mapping,
  title={Mapping LiNbO 3 phonon-polariton nonlinearities with 2D THz-THz-Raman spectroscopy},
  author={Lin, Haw-Wei and Mead, Griffin and Blake, Geoffrey A},
  journal={Physical Review Letters},
  volume={129},
  number={20},
  pages={207401},
  year={2022},
  publisher={APS}
}

@article{delaire2011giant,
  title={Giant anharmonic phonon scattering in PbTe},
  author={Delaire, Olivier and Ma, Jie and Marty, Karol and May, Andrew F and McGuire, Michael A and Du, Mao-Hua and Singh, David J and Podlesnyak, A and Ehlers, G and Lumsden, MD and others},
  journal={Nature materials},
  volume={10},
  number={8},
  pages={614--619},
  year={2011},
  publisher={Nature Publishing Group UK London}
}

@article{knoop2023anharmonicity,
  title={Anharmonicity in thermal insulators: An analysis from first principles},
  author={Knoop, Florian and Purcell, Thomas AR and Scheffler, Matthias and Carbogno, Christian},
  journal={Physical Review Letters},
  volume={130},
  number={23},
  pages={236301},
  year={2023},
  publisher={APS}
}

@article{wallace1965thermal,
  title={Thermal expansion and other anharmonic properties of crystals},
  author={Wallace, Duane C},
  journal={Physical Review},
  volume={139},
  number={3A},
  pages={A877},
  year={1965},
  publisher={APS}
}

@article{budai2014metallization,
  title={Metallization of vanadium dioxide driven by large phonon entropy},
  author={Budai, John D and Hong, Jiawang and Manley, Michael E and Specht, Eliot D and Li, Chen W and Tischler, Jonathan Z and Abernathy, Douglas L and Said, Ayman H and Leu, Bogdan M and Boatner, Lynn A and others},
  journal={Nature},
  volume={515},
  number={7528},
  pages={535--539},
  year={2014},
  publisher={Nature Publishing Group UK London}
}

@article{brebrick1963anomalous,
  title={Anomalous thermoelectric power as evidence for two-valence bands in SnTe},
  author={Brebrick, RF and Strauss, AJ},
  journal={Physical Review},
  volume={131},
  number={1},
  pages={104},
  year={1963},
  publisher={APS}
}

@article{hsieh2012topological,
  title={Topological crystalline insulators in the SnTe material class},
  author={Hsieh, Timothy H and Lin, Hsin and Liu, Junwei and Duan, Wenhui and Bansil, Arun and Fu, Liang},
  journal={Nature communications},
  volume={3},
  number={1},
  pages={982},
  year={2012},
  publisher={Nature Publishing Group UK London}
}

@article{su2023orietation,
  title={Orietation-controlled synthesis and Raman study of 2D SnTe},
  author={Su, Yanfei and Ding, Chuyun and Yao, Yuyu and Fu, Rao and Xue, Mengfei and Liu, Xiaolin and Lin, Jia and Wang, Feng and Zhan, Xueying and Wang, Zhenxing},
  journal={Nanotechnology},
  volume={34},
  number={50},
  pages={505206},
  year={2023},
  publisher={IOP Publishing}
}

@article{kannaujiya2020growth,
  title={Growth and characterizations of tin telluride (SnTe) single crystals},
  author={Kannaujiya, Rohitkumar M and Khimani, Ankurkumar J and Chaki, Sunil H and Chauhan, Sanjaysinh M and Hirpara, Anilkumar B and Deshpande, MP},
  journal={The European Physical Journal Plus},
  volume={135},
  pages={1--12},
  year={2020},
  publisher={Springer}
}

@inproceedings{vidal1992displacive,
  title={Displacive Excitation of Coherent Phonons},
  author={Vidal, J and Cheng, TK and Zeiger, HJ and Ippen, EP and Dresselhaus, G and Dresselhaus, MS},
  booktitle={International Conference on Ultrafast Phenomena},
  pages={ThC2},
  year={1992},
  organization={Optica Publishing Group}
}

@article{sasaki2018coherent,
  title={Coherent control theory and experiment of optical phonons in diamond},
  author={Sasaki, Hiroya and Tanaka, Riho and Okano, Yasuaki and Minami, Fujio and Kayanuma, Yosuke and Shikano, Yutaka and Nakamura, Kazutaka G},
  journal={Scientific Reports},
  volume={8},
  number={1},
  pages={9609},
  year={2018},
  publisher={Nature Publishing Group UK London}
}

@article{de2022decoupling,
  title={Decoupling of static and dynamic criticality in a driven Mott insulator},
  author={de la Torre, Alberto and Seyler, Kyle L and Buchhold, Michael and Baum, Yuval and Zhang, Gufeng and Laurita, Nicholas J and Harter, John W and Zhao, Liuyan and Phinney, Isabelle and Chen, Xiang and others},
  journal={Communications Physics},
  volume={5},
  number={1},
  pages={35},
  year={2022},
  publisher={Nature Publishing Group UK London}
}

@article{hase2002dynamics,
  title={Dynamics of coherent anharmonic phonons in bismuth using high density photoexcitation},
  author={Hase, Muneaki and Kitajima, Masahiro and Nakashima, Shin-ichi and Mizoguchi, Kohji},
  journal={Physical review letters},
  volume={88},
  number={6},
  pages={067401},
  year={2002},
  publisher={APS}
}

@article{fritz2007ultrafast,
  title={Ultrafast bond softening in bismuth: Mapping a solid's interatomic potential with x-rays},
  author={Fritz, David M and Reis, DA and Adams, B and Akre, RA and Arthur, J and Blome, Christian and Bucksbaum, PH and Cavalieri, Adrian L and Engemann, S and Fahy, S and others},
  journal={Science},
  volume={315},
  number={5812},
  pages={633--636},
  year={2007},
  publisher={American Association for the Advancement of Science}
}

@article{huang2022observation,
  title={Observation of a novel lattice instability in ultrafast photoexcited SnSe},
  author={Huang, Yijing and Yang, Shan and Teitelbaum, Samuel and De la Pe{\~n}a, Gilberto and Sato, Takahiro and Chollet, Matthieu and Zhu, Diling and Niedziela, Jennifer L and Bansal, Dipanshu and May, Andrew F and others},
  journal={Physical Review X},
  volume={12},
  number={1},
  pages={011029},
  year={2022},
  publisher={APS}
}

@article{cheng2017coherent,
  title={Coherent control of optical phonons in bismuth},
  author={Cheng, Yu-Hsiang and Gao, Frank Y and Teitelbaum, Samuel W and Nelson, Keith A},
  journal={Physical Review B},
  volume={96},
  number={13},
  pages={134302},
  year={2017},
  publisher={APS}
}

@article{ito2020observation,
  title={Observation of unoccupied states of SnTe (111) using pump-probe ARPES measurement},
  author={Ito, Hiroshi and Otaki, Yusuke and Tomohiro, Yuta and Ishida, Yukiaki and Akiyama, Ryota and Kimura, Akio and Shin, Shik and Kuroda, Shinji},
  journal={Physical Review Research},
  volume={2},
  number={4},
  pages={043120},
  year={2020},
  publisher={APS}
}

@article{horstmann2020coherent,
  title={Coherent control of a surface structural phase transition},
  author={Horstmann, Jan Gerrit and B{\"o}ckmann, Hannes and Wit, Bareld and Kurtz, Felix and Storeck, Gero and Ropers, Claus},
  journal={Nature},
  volume={583},
  number={7815},
  pages={232--236},
  year={2020},
  publisher={Nature Publishing Group UK London}
}

@article{maklar2023coherent,
  title={Coherent light control of a metastable hidden state},
  author={Maklar, Julian and Sarkar, Jit and Dong, Shuo and Gerasimenko, Yaroslav A and Pincelli, Tommaso and Beaulieu, Samuel and Kirchmann, Patrick S and Sobota, Jonathan A and Yang, Shuolong and Leuenberger, Dominik and others},
  journal={Science Advances},
  volume={9},
  number={47},
  pages={eadi4661},
  year={2023},
  publisher={American Association for the Advancement of Science}
}

@article{johnson2024all,
  title={All-optical seeding of a light-induced phase transition with correlated disorder},
  author={Johnson, Allan S and Pastor, Ernest and Batlle-Porro, Sergi and Benzidi, Hind and Katayama, Tetsuo and de la Pe{\~n}a Mu{\~n}oz, Gilberto A and Krapivin, Viktor and Kim, Sunam and L{\'o}pez, N{\'u}ria and Trigo, Mariano and others},
  journal={Nature Physics},
  pages={1--6},
  year={2024},
  publisher={Nature Publishing Group UK London}
}

@article{murray2005effect,
  title={Effect of lattice anharmonicity on high-amplitude phonon dynamics in photoexcited bismuth},
  author={Murray, ED and Fritz, DM and Wahlstrand, JK and Fahy, S and Reis, DA},
  journal={Physical Review B—Condensed Matter and Materials Physics},
  volume={72},
  number={6},
  pages={060301},
  year={2005},
  publisher={APS}
}

@article{maehrlein2017terahertz,
  title={Terahertz sum-frequency excitation of a Raman-active phonon},
  author={Maehrlein, Sebastian and Paarmann, Alexander and Wolf, Martin and Kampfrath, Tobias},
  journal={Physical review letters},
  volume={119},
  number={12},
  pages={127402},
  year={2017},
  publisher={APS}
}

@article{katayama2012ferroelectric,
  title={Ferroelectric soft mode in a SrTiO 3 thin film impulsively driven to the anharmonic regime using intense picosecond terahertz pulses},
  author={Katayama, I and Aoki, H and Takeda, J and Shimosato, H and Ashida, M and Kinjo, R and Kawayama, I and Tonouchi, M and Nagai, M and Tanaka, K},
  journal={Physical review letters},
  volume={108},
  number={9},
  pages={097401},
  year={2012},
  publisher={APS}
}

@article{von2018probing,
  title={Probing the interatomic potential of solids with strong-field nonlinear phononics},
  author={von Hoegen, Alexander and Mankowsky, Roman and Fechner, Michael and F{\"o}rst, Michael and Cavalleri, Andrea},
  journal={Nature},
  volume={555},
  number={7694},
  pages={79--82},
  year={2018},
  publisher={Nature Publishing Group UK London}
}

@article{melnikov2023anharmonic,
  title={Anharmonic coherent dynamics of the soft phonon mode of a PbTe crystal},
  author={Melnikov, AA and Selivanov, Yu G and Chekalin, SV},
  journal={Physical Review B},
  volume={108},
  number={22},
  pages={224309},
  year={2023},
  publisher={APS}
}

@article{hu2018femtosecond,
  title={Femtosecond study of A1g phonons in the strong 3D topological insulators: From pump-probe to coherent control},
  author={Hu, Jianbo and Igarashi, Kyushiro and Sasagawa, Takao and Nakamura, Kazutaka G and Misochko, Oleg V},
  journal={Applied Physics Letters},
  volume={112},
  number={3},
  year={2018},
  publisher={AIP Publishing}
}

@article{toberer2011phonon,
  title={Phonon engineering through crystal chemistry},
  author={Toberer, Eric S and Zevalkink, Alex and Snyder, G Jeffrey},
  journal={Journal of Materials Chemistry},
  volume={21},
  number={40},
  pages={15843--15852},
  year={2011},
  publisher={Royal Society of Chemistry}
}

@article{Suzuki_1995,
doi = {10.1143/JJAP.34.5977},
url = {https://dx.doi.org/10.1143/JJAP.34.5977},
year = {1995},
month = {nov},
publisher = {},
volume = {34},
number = {11R},
pages = {5977},
author = {Suzuki, Norihiro and Sadao Adachi, Sadao Adachi},
title = {Optical Properties of SnTe},
journal = {Japanese Journal of Applied Physics}
}

@article{beattie1969temperature,
  title={Temperature dependence of the elastic constants of tin telluride},
  author={Beattie, AG},
  journal={Journal of Applied Physics},
  volume={40},
  number={12},
  pages={4818--4821},
  year={1969},
  publisher={American Institute of Physics}
}

@Article{C4CP02091J,
author ="Zhou, Min and Gibbs, Zachary M. and Wang, Heng and Han, Yemao and Xin, Caini and Li, Laifeng and Snyder, G. Jeffrey",
title  ="Optimization of thermoelectric efficiency in SnTe: the case for the light band",
journal  ="Phys. Chem. Chem. Phys.",
year  ="2014",
volume  ="16",
issue  ="38",
pages  ="20741-20748",
publisher  ="The Royal Society of Chemistry",
doi  ="10.1039/C4CP02091J",
url  ="http://dx.doi.org/10.1039/C4CP02091J"}

@article{damon1966thermal,
  title={Thermal Conductivity of SnTe between 100 and 500 K},
  author={Damon, DH},
  journal={Journal of Applied Physics},
  volume={37},
  number={8},
  pages={3181--3190},
  year={1966},
  publisher={American Institute of Physics}
}

@article{he2016coherent,
  title={Coherent optical phonon oscillation and possible electronic softening in WTe2 crystals},
  author={He, Bin and Zhang, Chunfeng and Zhu, Weida and Li, Yufeng and Liu, Shenghua and Zhu, Xiyu and Wu, Xuewei and Wang, Xiaoyong and Wen, Hai-hu and Xiao, Min},
  journal={Scientific Reports},
  volume={6},
  number={1},
  pages={30487},
  year={2016},
  publisher={Nature Publishing Group UK London}
}

@article{xu2025time,
  title={Time-domain study of coupled collective excitations in quantum materials},
  author={Xu, Chenhang and Zong, Alfred},
  journal={npj Quantum Materials},
  volume={10},
  number={1},
  pages={21},
  year={2025},
  publisher={Nature Publishing Group UK London}
}

@article{giorgianni2022terahertz,
  title={Terahertz displacive excitation of a coherent Raman-active phonon in V2O3},
  author={Giorgianni, Flavio and Udina, Mattia and Cea, Tommaso and Paris, Eugenio and Caputo, Marco and Radovic, Milan and Boie, Larissa and Sakai, Joe and Schneider, Christof W and Johnson, Steven Lee},
  journal={Communications Physics},
  volume={5},
  number={1},
  pages={103},
  year={2022},
  publisher={Nature Publishing Group UK London}
}

@article{barrios2014chemically,
  title={Chemically deposited SnSe thin films: thermal stability and solar cell application},
  author={Barrios-Salgado, Enue and Nair, MTS and Nair, PK},
  journal={ECS Journal of Solid State Science and Technology},
  volume={3},
  number={8},
  pages={Q169},
  year={2014},
  publisher={IOP Publishing}
}

@article{frohlich1937theory,
  title={Theory of electrical breakdown in ionic crystals},
  author={Fr{\"o}hlich, Herbert},
  journal={Proceedings of the Royal Society of London. Series A-Mathematical and Physical Sciences},
  volume={160},
  number={901},
  pages={230--241},
  year={1937},
  publisher={The Royal Society London}
}
